**PAPER**

# Confirmation of a non-transiting planet in the habitable zone of the nearby M dwarf L 98-59

Paul I. Schwarz*[1] | Stefan Dreizler[1] | René Heller[2]

[1]Institute for Astrophysics and Geophysics (IAG), Georg-August-Universität, Lower Saxony, Germany
[2]Solar and Stellar Interiors, Max Plank Institute for Solar System Research, Lower Saxony, Germany

**Correspondence**
*Paul Ibrahim Schwarz, Email: paul.schwarz@uni-goettingen.de

**Present Address**
Paul I. Schwarz, Ludwig-Prandtl-Straße 18, 37077 Göttingen, Germany

Only 40 exoplanetary systems with five or more planets are currently known. These systems are crucial for our understanding of planet formation and planet-planet interaction. The M dwarf L 98-59 has previously been found to show evidence of five planets, three of which are transiting. Our aim is to confirm the fifth planet in this system and to refine the system characteristics namely minimum masses, radii and the orbital parameters of the planets around L 98-59. We reanalysed RV and activity data from HARPS and ESPRESSO alongside TESS and HST transit data using a joint model. The parameter space was sampled using the dynesty nested sampler. We confirm the previously known fifth planet in the system's habitable zone with an orbital period of 23.07 ± 0.08 d, a minimum mass of $3.0 \pm 0.5\,M_\oplus$ and an effective temperature of 289 K. We find an additional planet candidate in the RV data with an orbital period of $1.7361^{+0.0007}_{-0.0008}$ d and a minimum mass of $0.58 \pm 0.12\,M_\oplus$. This candidate (L 98-59.06) has a statistical significance between $2.9\,\sigma$ and $4.2\,\sigma$, details depending on the modelling of stellar variability. Moreover, we present evidence for a stellar rotation period of 76 ± 4 d.

## 1 | INTRODUCTION

The M dwarf star L 98-59 (also known as TOI-175, TIC 307210830 and others) has been known to host four confirmed planets, referred to as L 98-59 b, c, d and e (Demangeon et al. 2021b, hereafter D21), the inner three of which are transiting (Kostov et al. 2019, hereafter K19). High-resolution radial velocity (RV) data suggests the presence of a fifth, non-transiting planet (D21), which would be the outermost known planet of this system and in the stellar habitable zone (Kasting, Whitmire, & Reynolds, 1993). Throughout this paper, we refer to this outermost object as L 98-59 f, which was proposed as a planetary candidate with an orbital period of $23.15^{+0.60}_{-0.17}$ d and a minimum mass of $2.46^{+0.66}_{-0.82}\,M_\oplus$ by D21.

The orbital and planetary parameters of the system have been studied in great detail before (Demangeon et al., 2021b; Howard et al., 2021; Stassun et al., 2019)and RV data. This system is thus known as one of the most well-constrained multi-planet systems with mass uncertainties below 25 % and radius uncertainties below 8 % (Luque & Pallé, 2022). The stellar parameters are summarised in Table 1 .

Its three transiting planets L 98-59 b, c and d have recently been observed for atmospheric characterisation (Zhou, Ma, Wang, & Zhu, 2022; Zhou, Ma, Wang, & Zhu, 2023). And ongoing observations with the James Webb Space Telescope (JWST) could soon allow unprecedented insights into the atmospherical composition of Earth-sized, nearby planets.Seligman et al. (2024) argue for the detectability of vulcanism on all three transiting planets using five to ten transits with JWST. Fromont et al. (2024) propose the detectability of atmospheric oxygen as a remnant of water evaporation and hydrogen escape due to the high XUV irradiation.

L 98-59 is also listed for investigation with the Large Interferometer For Exoplanets (LIFE) mission (Carrión-González et al., 2023) and in the input catalogue of the Habitable World Observatory (Tuchow, Stark, & Mamajek, 2024) due to its proximity to the solar system at a distance of only 10.6 pc. L 98-59 is also close to the edge of the first long-observation phase field (LOPS2) of the PLATO Mission and could therefore potentially be covered with high-quality space-based light curves for several years (Nascimbeni et al., 2025).

A peculiar challenge of L 98-59 and its planets lies in the non-negligible stellar activity, which complicates the unequivocal separation of the stellar and planetary nature of any signals. Two stellar rotation periods have previously been discussed in the literature. A study of the activity indicators of the HARPS data and of the correlation of the rotation period with the $R'_{HK}$ index suggests that the rotation period is around $78 \pm 13$ d (Astudillo-Defru et al., 2017; Cloutier et al., 2019). In a second analysis by C19 the authors include a Gaussian Process (GP) model to fit the stellar activity in the RV data and find a stellar rotation period of $51.4^{+6.5}_{-24.3}$ d. Howard et al. (2021) used TESS light curves and observations from Evryscope (Law et al., 2015; Ratzloff et al., 2019) and find a rotation period of $39.6 \pm 2.2$ d, which coincides with half of the previously proposed rotation period of about 78 d. From the analysis of the stellar activity indicators extracted from ESPRESSO spectra D21 derived a stellar rotation period of $80.9^{+5.0}_{-5.3}$ d. However, in their joint analysis of the RV and transit data with a GP for the RVs these authors find a rotation period of $33^{+43}_{-19}$ d. So the case of the stellar rotation period remains subject to debate and we aim to settle it in this paper.

Given the high precision in the masses, radii, and ephemerides required for further investigations of the planetary compositions and atmospheres, we aim to improve the determination planetary system's properties to better characterise its architecture. For our study, we use twelve new sectors of TESS data in a combined model for the transit and RV data that also takes the activity indicators into account.

## 2 | OBSERVATIONS

### 2.1 | Stellar spectroscopy

#### 2.1.1 | HARPS

To confirm the planetary candidate L 98-59 f in the HARPS (Mayor et al., 2003) data, we use the publicly available spectra of Cloutier et al. (2019) (hereafter C19) from the ESO science archive. We then apply an independent extraction via the SERVAL package (Zechmeister et al., 2018). SERVAL uses least-squares fitting of co-added spectra to extract RVs and activity indicators.

**TABLE 1** Stellar parameters of L 98-59. Howard et al. (2021), here H21.

| | value | source |
|---|---|---|
| RA [hh:mm:ss.ss] | 08:18:07.89 | GAIA-DR2 |
| DEC [dd:mm:ss.ss] | -68:18:52.08 | GAIA-DR2 |
| Sp. Type | M3V | K19 |
| $M_\star$ [$M_\odot$] | $0.29 \pm 0.02$ | TIC v8 |
| $R_\star$ [$R_\odot$] | $0.314 \pm 0.009$ | TIC v8 |
| $T_{\rm eff}$ [K] | $3415 \pm 135$ | D21 |
| $P_{\rm rot}$ [d] | $76 \pm 4$ | This Work |
| distance [pc] | $10.6194 \pm 0.0032$ | TIC v8 |
| $\log g$ [cm s$^{-2}$] | $4.86 \pm 0.13$ | D21 |
| [Fe/H] [dex] | $-0.46 \pm 0.26$ | D21 |
| age [Gyr] | 0.8 to 10 | D21 |

The observations consist of 166 spectra taken between October 17, 2018 (2 458 408.5 BJD) and April 28, 2019 (2 458 601.5 BJD)[1]. As discussed in C19 the wavelengths were not simultaneously calibrated. Of the total 166 spectra, 141 were taken with an exposure time of 900 s, while the others were observed with exposures ranging from 500 s to 1800 s.

The extracted RV data show a median precision of 2.08 m s$^{-1}$. After an iterative $4\sigma$ clipping, six radial velocity data points were excluded (see Table 2 ). We compared these outliers to outliers found in previous analyses (C19; D21) and noticed that these are not identical. We attribute this to the different extraction methods.

#### 2.1.2 | ESPRESSO

Between November 14, 2018 (2 458 436.5 BJD) and March 4, 2020 (2 458 912.5 BJD) 66 spectra were taken by D21[2], each with 900 s exposure. We used the RV and activity indicators as published by Demangeon et al. (2021a), who used the pipeline of the ESPRESSO (Pepe et al., 2021) data reduction software to extract the values. We did not redo extraction with SERVAL. Following D21, we also exclude 2 458 645.496 BJD, 2 458 924.639 BJD and 2 458 924.645 BJD due to large uncertainties.

[1] Observations under program IDs 198.C-0838(A), 1102.C-0339(A) and 0102.C-0525

[2] Observations under program IDs 1102.C-0744, 1102.C-0958 and 1104.C-0350

**TABLE 2** Discarded data points from a $4\sigma$ clipping of the values and exclusion of data points with large errors.

| time [BJD − 2458354] | RV [m s$^{-1}$] | error [m s$^{-1}$] |
|---|---|---|
| 78.82823 | 3.46 | 5.45 |
| 135.83276 | -1.91 | 4.20 |
| 149.79586 | -18.11 | 1.78 |
| 196.72648 | 4.65 | 8.23 |
| 243.55736 | -0.20 | 5.73 |
| 243.57330 | -3.58 | 4.05 |

We separated the RV data from ESPRESSO into two sets: the pre- and post-fibre upgrade parts. This enabled us to fit them with individual offsets, that might occur due to shifts in the extraction baseline.

## 2.2 | Stellar photometry

We also used transit observations from TESS and HST of L 98-59 b. The TESS observations were taken in the short cadence mode (2 min) and correspond to almost 5 years of in-homogeneously sampled transit light curves covered by 21 sectors of TESS observations (2, 5, 8, 9, 10, 11, 12, 28, 29, 32, 35, 36, 37, 38, 39, 61, 62, 63, 64, 65 and 69) between August 23, 2018 and September 18, 2023. To our knowledge, the twelve sectors 32 to 69 have not previously been used in the literature.

We used the Pre-search Data Conditioning Simple Aperture Photometry (PDCSAP) flux, which is already corrected for instrumental noise. To account for stellar variability, we used an additional r-spline detrending as implemented in the `WOTAN` package (Hippke, David, Mulders, & Heller, 2019), see appendix APPENDIX A:.

Moreover, we used five white light transit observations of the innermost planet L 98-59 b from HST's Wide Field Camera 3 provided by (Zhou et al., 2022). These data have not been used in a combined analysis for the orbital and planetary parameters before. We did not subject these HST transits to any additional detrending. HST observed the system from February 9, 2020 to February 24, 2021.

# 3 | MODELS AND METHODS

We use a Python package called `eff` (`Exoplanet Flexi-Fit`, Dreizler (2021)). It is designed for combined fits of RV and transit data of exoplanet systems and it offers different orbital approximations such as circular orbits, Keplerian orbits, or full $n$-body integration. The package `eff` includes nested sampling for estimating Bayesian posteriors with the `dynesty`. It also supports modelling of stellar activity in the RV data with Gaussian Processes (GP) using `celerite`, transit and RV modelling as well as outputs using GLS periodograms, corner plots (Foreman-Mackey, 2016) and phase folded RVs and transit light curves.

In our modelling of the planetary system, we investigate three possible scenarios for planetary multiplicity. First, examine a four planet model, which consists of the three transiting planets with periods of 2.25 d (planet b), 3.69 d (planet c) and 7.45 d (planet d) and the non-transiting planet with an orbital period of 12.8 d (planet e). For the five-planet model we include the non-transiting candidate from D21 with an orbital period of 23.1 d (planet f). In the six-planet model we additionally search for an additional, non-transiting candidate with an orbital period near 1.74 d (planet candidate L 98-59.06).

## 3.1 | Modelling of the combined RV and transit data

For modelling the transit light curves we used the analytic transit model with quadratic limb-darkening (Mandel & Agol, 2002) in the implementation from Crossfield (2011).

The posterior sampling is implemented with the dynamic nested sampling algorithm `dynesty` (Speagle (2020), v.2.1.1; Koposov et al. (2023)). Nested sampling algorithms such as `dynesty` estimate the marginal likelihood or Bayesian evidence ($\mathcal{Z}$) alongside the posterior by integrating the priors of iso-likelihood regions. `dynesty` provides a robust estimation of the statistical and sampling uncertainties. We stop the sampling when the contribution of the remaining prior volume to the final evidence of $\Delta \ln \mathcal{Z} \leq 0.1$ is reached.

The GP modelling is implemented using the package `celerite` (Foreman-Mackey, Agol, Ambikasaran, & Angus, 2017). `Celerite` employs GP modelling with an improved algorithm using exponentials in the kernel. When inverting the covariance matrix to calculate marginalised likelihoods, `celerite` exploits the semiseparable structure of these matrices to speed up the calculations for which a generalised solver was derived by Ambikasaran (2014). This enables `celerite` to decrease the calculation time to scale linearly with the number of data points $N$.

Our model consists of a common set of orbital and planetary parameters as well as the parameters for the GP model to account for the stellar activity in the RV data. For the GP model we use different kernel functions (Sec. 3.2). This was not necessary for the transit data, due to a better distinction of the rotational signature in the transits, and a better detrending (see Sec. 2.2).

The model fit to the combined data from TESS, HST, HARPS and ESPRESSO is built upon a common set of orbital and planetary parameters from which the light curve models and the radial velocities are extracted. To minimise any systematic bias on the planetary and orbital parameters, we set most of the prior distributions to be uniform ($\mathcal{U}$). They had wide boundaries derived from preliminary runs which were consistent with the posteriors from D21. For the $h$ and $k$ parameters

$$h = \sqrt{e}\sin(\omega), \quad k = \sqrt{e}\cos(\omega), \tag{1}$$

where $e$ is the eccentricity and $\omega$ the argument of periastron, we chose truncated normal distributions (trunc$\mathcal{N}$) as priors, with boundaries ensuring eccentricities below 1. The period priors in our search for an additional, sixth planet were motivated by the period ratio model posteriors from Dietrich (2024). These authors predict another planet near 5 d for which we do not find any evidence.

## 3.2 | Correlated noise kernels

To account for the influence of the stellar variability on the RVs, we use a GP model with three different kernels to capture stellar activity. Our first choice is a single stochastically damped simple harmonic oscillator (sSHO) from the `celerite` package (Foreman-Mackey et al., 2017). Second, we include a double SHO (dSHO) to better account for aliasing (Foreman-Mackey, 2018). This kernel is parameterised with the lead period $P$, a mean quality factor $Q_0$ combined with a deviation term $dQ$ as well as a factor between the SHO kernels $f$ and an amplitude scale $\sigma$:

$$\begin{aligned} Q_1 &= 0.5 + Q_0 + dQ, & Q_2 &= 0.5 + Q_0 \\ \omega_1 &= \frac{4\pi Q_1}{P\sqrt{4Q_1^2 - 1}}, & \omega_2 &= \frac{8\pi Q_2}{P\sqrt{4Q_2^2 - 1}} \\ S_1 &= \frac{1}{1+f}\frac{\sigma^2}{\omega_1 Q_1}, & S_2 &= \frac{f}{1+f}\frac{\sigma^2}{\omega_2 Q_2} \end{aligned} \tag{2}$$

And third, we include a coupled double SHO (cdSHO), which is similar to the dSHO but uses a coupled quality factor,

$$Q_1 = 0.5 + Q_0, \quad Q_2 = 2Q_1 \tag{3}$$

term to ensure the same damping timescale for both oscillators,

The prior distributions are set in the parameter space of the input, which are translated to the `celerite` sampling parameters (compare Table 3 ).

## 3.3 | Activity Indicators

The activity of the host star was assessed by examining multiple activity indicators. We use the activity indicators from the `SERVAL` data extraction of the HARPS data and the ones published by D21.

- Line indices such as the depths of the $H_\alpha$ line and the sodium doublet (NaD, Díaz, Cincunegui, and Mauas 2007), which are available for both datasets
- Chromatic RV indeX (CRX, HARPS only), defined as the slope of a linear fit of the logarithmic wavelength dependence of the RVs (Zechmeister et al., 2018)
- Differential Line Width (dLW, HARPS only), measuring the scaled deviation of the second derivative of the (cubic B-spline) template from the measured spectra (Zechmeister et al., 2018)
- S-Index (ESPRESSO only), defined as the ratio of the Calcium II H&K band-passes and the continuum R&V band-passes (Vaughan, Preston, & Wilson, 1978).
- BISector (BIS, ESPRESSO only), defined as the difference in velocity of the upper and lower part in a mean spectral line profile in the Cross-Correlation Function (CCF) of all non-saturated spectral lines (Queloz et al., 2001)

For the examination of any periodic behavior of the activity indicators we utilise power spectra, specifically the Generalised Lomb-Scargle (GLS) periodogram (Zechmeister & Kürster, 2009). Figure 1 shows the GLS periodograms of these activity indicators of the HARPS (*left*) and ESPRESSO (pre-fibre upgrade *middle*, post-fibre upgrade *right*) data. As with the RV data we excluded some measurements of the activity data with the same $4\sigma$ clipping or with unreasonably large uncertainties.

## 3.4 | Additional algorithms

We use the `L1`-periodograms from Hara, Boué, Laskar, and Correia (2017) to assess the signatures present in the RV data as an independent reinforcement of our findings. The periodicities of unevenly sampled data sets are analysed by searching for representations with a restricted set of sinusoidal signals. This approach helps to minimise the number of peaks in the periodogram.

**TABLE 3** Kernel functions and parametrisations.

| abbreviation | input | sampling |
|---|---|---|
| sSHO | $\sigma$, $P$, $Q_0$ | $S_0$, $\omega$, $Q_0$ |
| dSHO | $\sigma$, $P$, $Q_0$, $dQ$, $f$ | $S_1$, $\omega_1$, $Q_1$, $S_2$, $Q_2$ |
| cdSHO | $\sigma$, $P$, $Q_0$, $f$ | $S_1$, $\omega_1$, $Q_1$, $S_2$ |

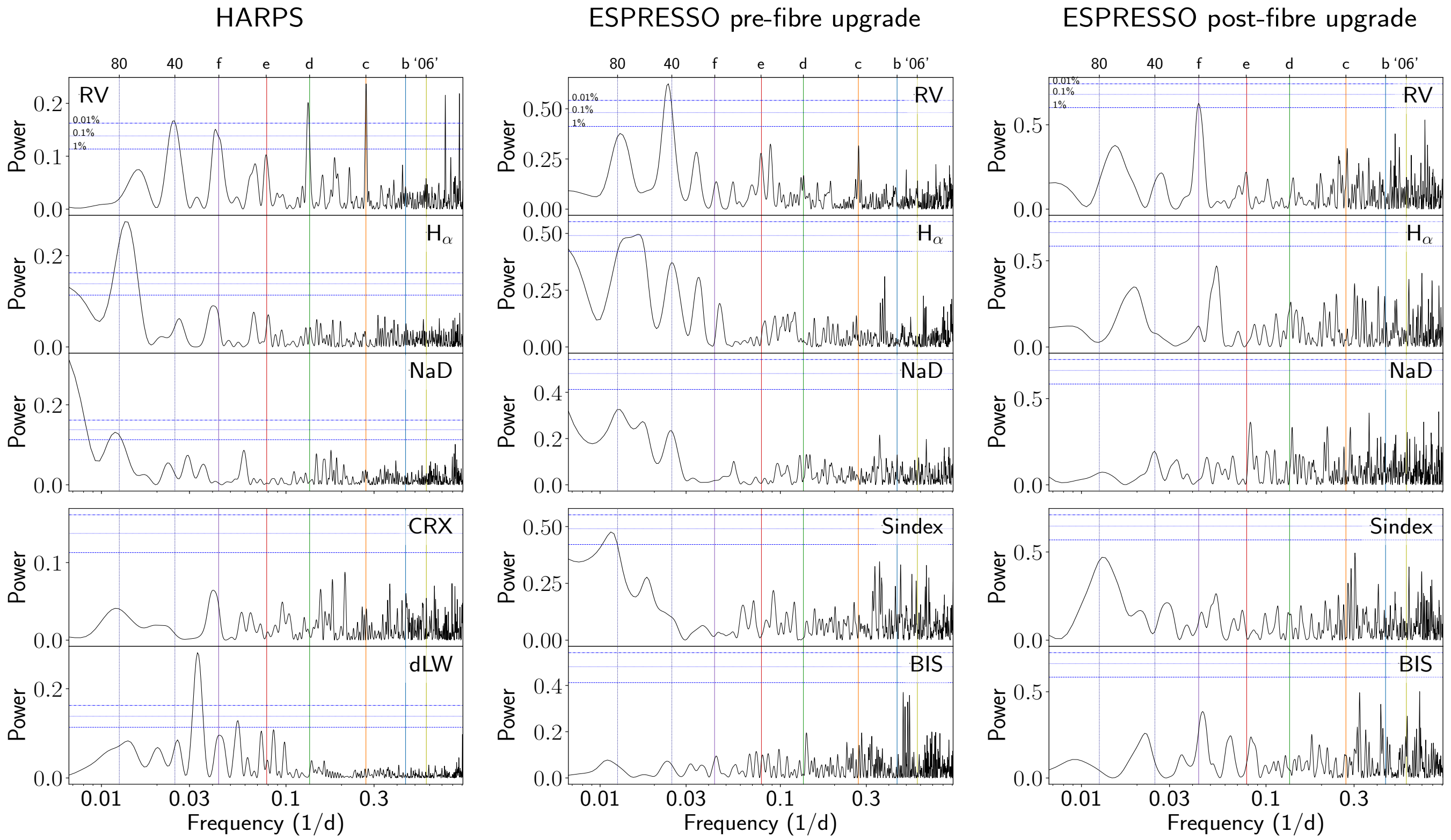

**FIGURE 1** Generalised Lomb-Scargle periodograms of the RVs (*top row*) and several stellar activity indicators. Vertical solid lines denote the orbital periods of the planets and the sixth candidate. Values along the ordinate correspond to the power normalised according to Zechmeister and Kürster (2009). Vertical dotted lines indicate supposed stellar activity. The horizontal lines indicate false alarm probabilities (FAP) of 1%, 0.1% and 0.01% from the bottom to top. *Left*: RVs and activity indicators from the `SERVAL` analysis of the HARPS spectra. *Middle*: RVs and activity indicators from D21 of the ESPRESSO spectra pre-fibre upgrade. *Right*: RVs and activity indicators from D21 of the ESPRESSO spectra post-fibre upgrade.

In our search for previously undetected transits we use the Transit Least Squares (`TLS`) Python package (Hippke & Heller, 2019). `TLS` is designed for the search of small planets and achieves a higher Signal Detection Efficiency (SDE) than the widely used box-fitting algorithms for transits (Heller et al., 2020; Heller, Hippke, & Rodenbeck, 2019; Heller, Rodenbeck, & Hippke, 2019). `TLS` searches for transit signatures in light curves by phase folding over different transit epochs, trial periods, and transit durations using astrophysically motivated transits profiles with stellar limb darkening.

## 4 | RESULTS

### 4.1 | Stellar activity

The analytically determined FAP levels from (Zechmeister & Kürster, 2009) are presented as horizontal blue lines in Fig. 1 and Fig. 2 (0.01 % dash-dotted, 0.1 % dotted and 1 % dashed).

The activity data used for this study are summarised in Fig. 1 . The $H_\alpha$ and NaD activity indicators can be directly compared between the HARPS and ESPRESSO data, while the CRX and dLW are only available from the SERVAL pipeline, while the S-index and BIS are available from the ESPRESSO data-reduction pipeline. Comparing the $H_\alpha$ and NaD with the RV data we find strong peaks towards higher periods. From previous analyses we know that the rotation period is either around 78 d or 39 d. In the HARPS $H_\alpha$ line index the FAP around 80 d exceeds a FAP of 0.01 %. This peak is visible in some of the other activity indicators as well, but not nearly as significant, mostly not even reaching a FAP of 1 %. Furthermore the HARPS dLW shows a strong peak exceeding 0.01 % FAP near 33 d which we cannot directly attribute to a known activity cycle of L 98-59, but could be an artifact of the second harmonic of the stellar rotation.

We test three different GP kernels (see Sec. 3.2), with different period priors, namely normal distributions ($\mathcal{N}$) around 39±5 d and 78±5 d as well as a wide uniform distribution ($\mathcal{U}$) from 25 d to 500 d. These priors are referred to as $\mathcal{N}(39, 5)$, $\mathcal{N}(78, 5)$ and $\mathcal{U}(25, 500)$ in Table 4 , respectively.

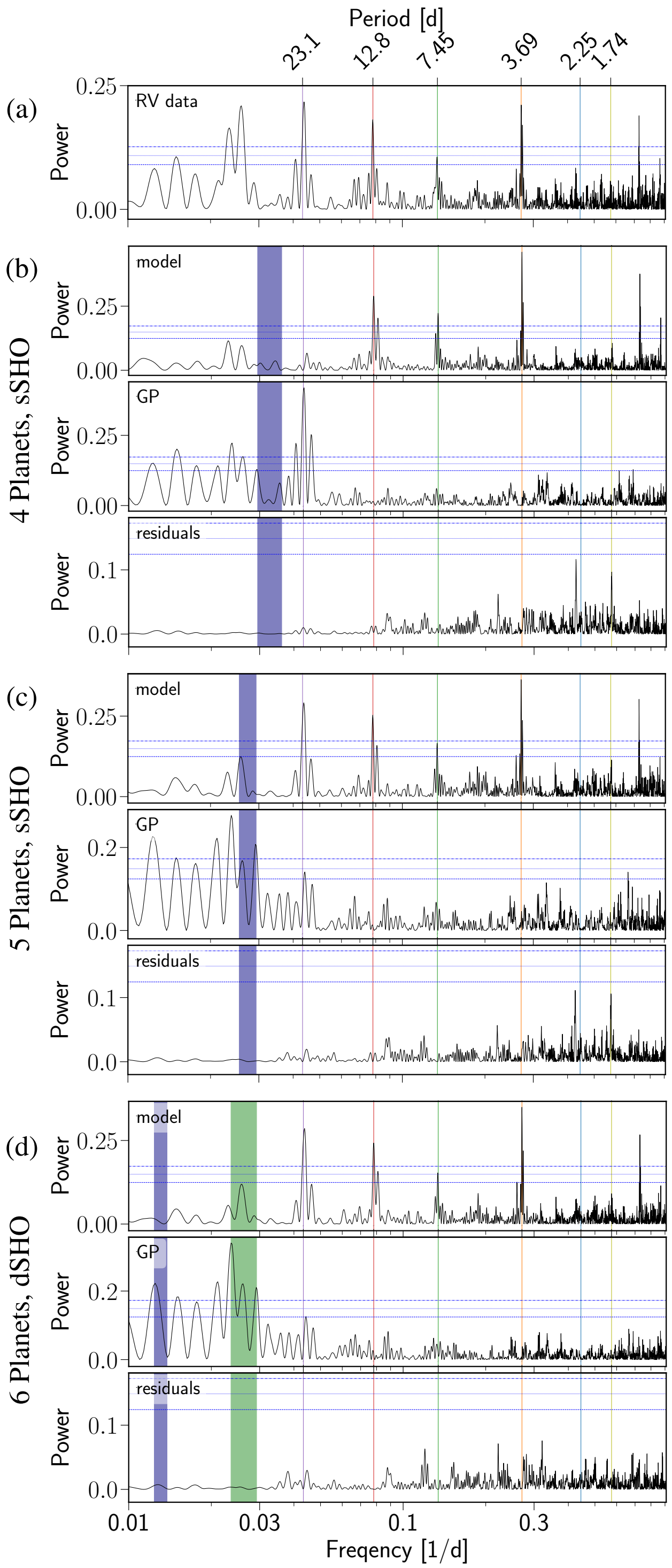

**FIGURE 2** *(a)* GLS periodograms for the combined RV data. *(b)* Panel triplet with the best fit results of the four planet model. *(c)* Five-planet model. *(d)* Six-planet model. The subpanels in *(b)*-*(d)* show the GLS periodograms of the best-fitting RV model, GLS periodograms of the GP, and GLS periodograms of the corresponding residuals.

Since the GP was used in addition to the planet models on the RV data, the models yield slightly different results for the stellar activity. The four planet model with the sSHO $\mathcal{U}(25, 500)$ GP results in the highest evidence, with a rotation period of $28.0^{+2.4}_{-1.7}$ d. As shown in Fig. 2 [3] the GP in the four planet model absorbs the signature of the fifth planet. For the five-planet model the sSHO $\mathcal{N}(39, 5)$ GP yields the highest evidence with a posterior stellar rotation period of $37.7^{+3.2}_{-3.2}$ d. In the six-planet model the highest evidence is achieved with the dSHO $\mathcal{N}(78, 5)$ GP. Here a stellar rotation period of $76.6^{+4.1}_{-4.2}$ d is preferred. The factor between the two SHOs, f is high ($18 \pm 10$) which indicates that the main power is in the second harmonic rather than the first (compare Eq. (2)). These results are present in both the log-likelihood and the Bayesian evidence (compare Table C1 ).

## 4.2 | Confirmation of the fifth planet

We show the GLS periodograms of the raw RV data, as well as the best models for the four, five and six-planet models in Fig. 2 . The coloured, vertical lines represent the position of the planet signals, labelled with their respective periods in days. The shaded regions indicate the best fitting GP period. Table 4 summarises the evidence and log-likelihood increase for the different noise models for a full table of the evidences and log-likelihoods see Table C1 .

With our analysis we find that the GPs in the five-planet model do not significantly reduce the RV amplitude of the fifth planet. As summarised in Table 4 the five planet models are at least moderately better than the four planet models when we exclude the sSHO with the wide priors, since there the signal of the fifth candidate from D21 is included in the GP (compare Fig. 2 *(b)* GP and residuals). For the best fitting five planet model (with the sSHO $\mathcal{N}(39, 5)$) we have a better fit by $3.6\sigma$ and a $\Delta \ln \mathcal{Z}$ of 7.9 corresponding to strong[4] evidence in favour of this planet. The model without a GP resulted in an even higher evidence increase ($\Delta \ln \mathcal{Z} = 11.3$). With an average $\Delta \ln \mathcal{Z} \sim +5.7$ we find strong evidence for the five planet model and are confident to confirm the planetary nature of the fifth planet.

[3] Here the vertical lines indicate the orbital periods of the planets and the shaded areas the $1\sigma$ region (blue) of the best fit stellar activity periodicity. The horizontal lines mark the FAP of 1%, 0.1% and 0.01% from the bottom to top.

[4] Trotta (2008) finds a $\Delta \ln \mathcal{Z} > 5$ as strong and $\Delta \ln \mathcal{Z} > 2.5$ as moderate evidence for models.

## 4.3 | Exploration of the additional signal

### 4.3.1 | Alias rejection

In the residuals of the best four- and five-planet models, we see strong peaks at 1.736 d and 2.34 d, which are 1 d aliases of each other: $P_{\rm alias} = \left((1\,{\rm d})^{-1} + (1.736\,{\rm d})^{-1}\right)^{-1} \approx 2.34\,{\rm d}$.

The signal with a period at 2.34 d, however, can be excluded from stability considerations. As a metric for this, we use the separation of two adjacent planets ($j$ and $j+1$) in units of mutual Hill radii $\Delta(R_H)$ (Gladman, 1993). This can be calculated as per

$$\Delta(R_H)_{j+1,j} = 2\frac{a_{j+1} - a_j}{a_{j+1} + a_j}\left(\frac{3M}{m_{j+1} + m_j}\right)^{1/3},$$

where $M$ is the stellar mass, $m_j$ and $m_{j+1}$ are the planetary masses, and $a_j$ and $a_{j+1}$ are the semi-major axes (Obertas, Van Laerhoven, & Tamayo, 2017). With the mass of the new planet candidate $m_{06} > 0$, we obtain $\Delta(R_H)_{06,\,\rm b} \leq 2.5$. From results by Weiss et al. (2018) (hereafter W18) we know, that long term stability can be estimated with this metric. For example only about 10 % of the *Kepler* multi planet systems have $\Delta(R_H) < 10$. Generally a separation of $\Delta(R_H) \geq 13$ can be assumed for long time stability due to the orbital crossing timescale (Dreizler et al., 2024). In conclusion, we consider a period of 2.34 d as unstable and, hence, reject it as a possible planetary period.

### 4.3.2 | Transit search

In our search for possible transits of L 98-59.06 we checked the residuals of the light curves for additional transits with TLS. We found no significant peaks in the SDE periodogram from 0.1 d to 2 d. Therefore we expand the five planet model with a sixth non-transiting planet.

### 4.3.3 | FAP estimation and injection retrieval

To assess the reliability of the candidate signature at 1.736 d we perform a bootstrapping using the residuals of the RV data after subtracting the five planet model. We randomly reassigned the measurements, including their error bars, to timestamps from the original data. Then we located the strongest power in the resulting GLS periodogram.

This process was repeated 10 000 times. In 0.46 % of these simulations, we found a power that was equal or larger than the power at 1.736 d that we found in the original RV data. This numerical FAP is significantly lower than the FAP of 5.3 % which we found in the residuals of the RV data for the five planet model. We are thus confident that this signal can be distinguished from white noise.

We also performed an injection-retrieval test with comparable RV amplitude and RV period in the residuals. This led to a reliability of $R = T_{\rm rec}/N_{\rm inj} = 90.4\,\%$, with $T_{\rm rec}$ as the number of true signals recovered and $N_{\rm inj}$ as the number of trials.

### 4.3.4 | Prior estimation from RV only fits

We also explore the effect of various period priors on the resulting statistical evidence of any possible RV signals from a sixth planet. We find that the best fit and the median of the period posteriors are all statistically compatible with 1.736 d within one standard deviation, independent of the prior. We test uniform priors on an evenly sampled frequency grid with varying boundaries: $\mathcal{U}(1.1\,{\rm d}, 1.9\,{\rm d})$ and $\mathcal{U}(1.55\,{\rm d}, 1.9\,{\rm d})$. We find that the log-likelihood increase from a five-planet model to a six-planet model does indeed favour the new sixth planet for all three priors that we tested.

Finally, we also test a shifted prior ($\mathcal{U}(1.6\,{\rm d}, 1.68\,{\rm d})$ to examine whether we could also detect a signal in the residuals of a period interval, in which we do not see a peak in the GLS periodograms. This experiment did not lead to any log-likelihood increases in comparison to the five planet model.

### 4.3.5 | Statistical evaluation of the planet candidate

We examine the model fits for the combined RV and transit data. The full comparison of all different models and GPs for the stellar activity estimation, are summarised in Table C1 . Table 4 provides an overview of the different models, grouped by GP kernel function. The columns for the $\Delta \log \mathcal{L}$, the $\Delta \ln \mathcal{Z}$ and the $\sigma$ show the differences in the respective metrics between models with different numbers of planets (see column #). The $\sigma$-level differences were computed according to Heller and Hippke (2024).

We find that the six-planet model exhibits a significantly higher $\ln \mathcal{Z}$ and $\log \mathcal{L}$ for all noise models. The six-planet model is always at least 2.9 σ better than the five planet model. Additionally, we find that the best six-planet model (six planets and a dSHO GP with a leading period of $76.6^{+4.1}_{-4.2}$ d) exhibits an evidence increase of $\Delta \ln \mathcal{Z} = 10$ and a log-likelihood increase of $\Delta \log \mathcal{L} = 12.6$ with respect to the best fitting five planet model (five planets and a sSHO GP with a rotation period of $37.7^{+3.2}_{-3.2}$ d).

The best fit results from our analysis are summarised in the following figures. Figure 3 shows the phase folded RV data and models for all planets and the planet candidate. Figure 4 shows the combined planet signals and the GP over the full time of observations as well as the data. In Fig. 5 we show the phase folded transits from the HST observations. The TESS data is summarised by selected transits (see Fig. 6 ). We selected overlapping transits to represent the quality of the fit.

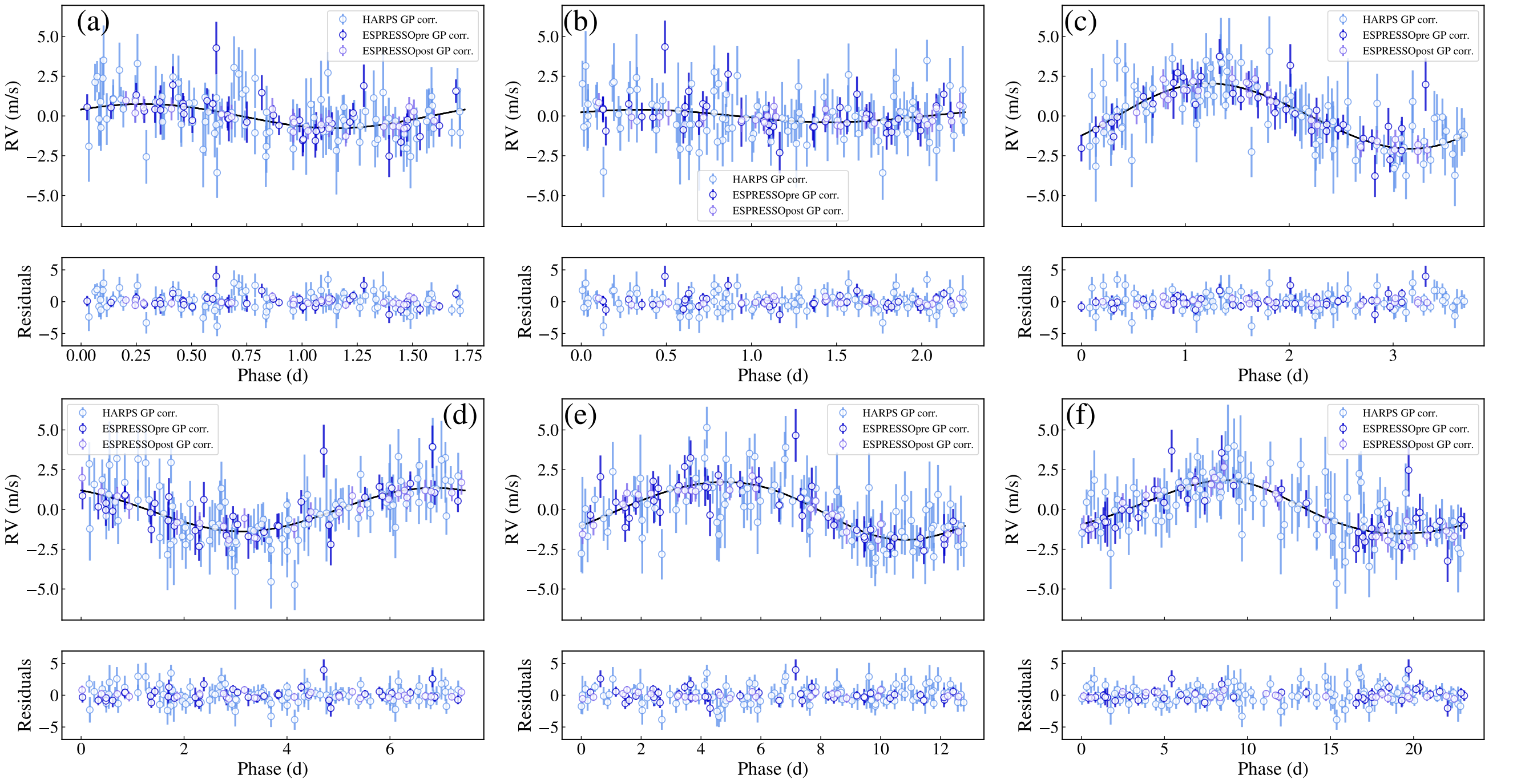

**FIGURE 3** Phase-folded radial velocities of the planets in the L 98-59 system. *(a)*: L 98-59.06. *(b)*: L 98-59 b. *(c)*: L 98-59 c. *(d)t*: L 98-59 d. *(e)*: L 98-59 e. *(f)*: L 98-59 f.

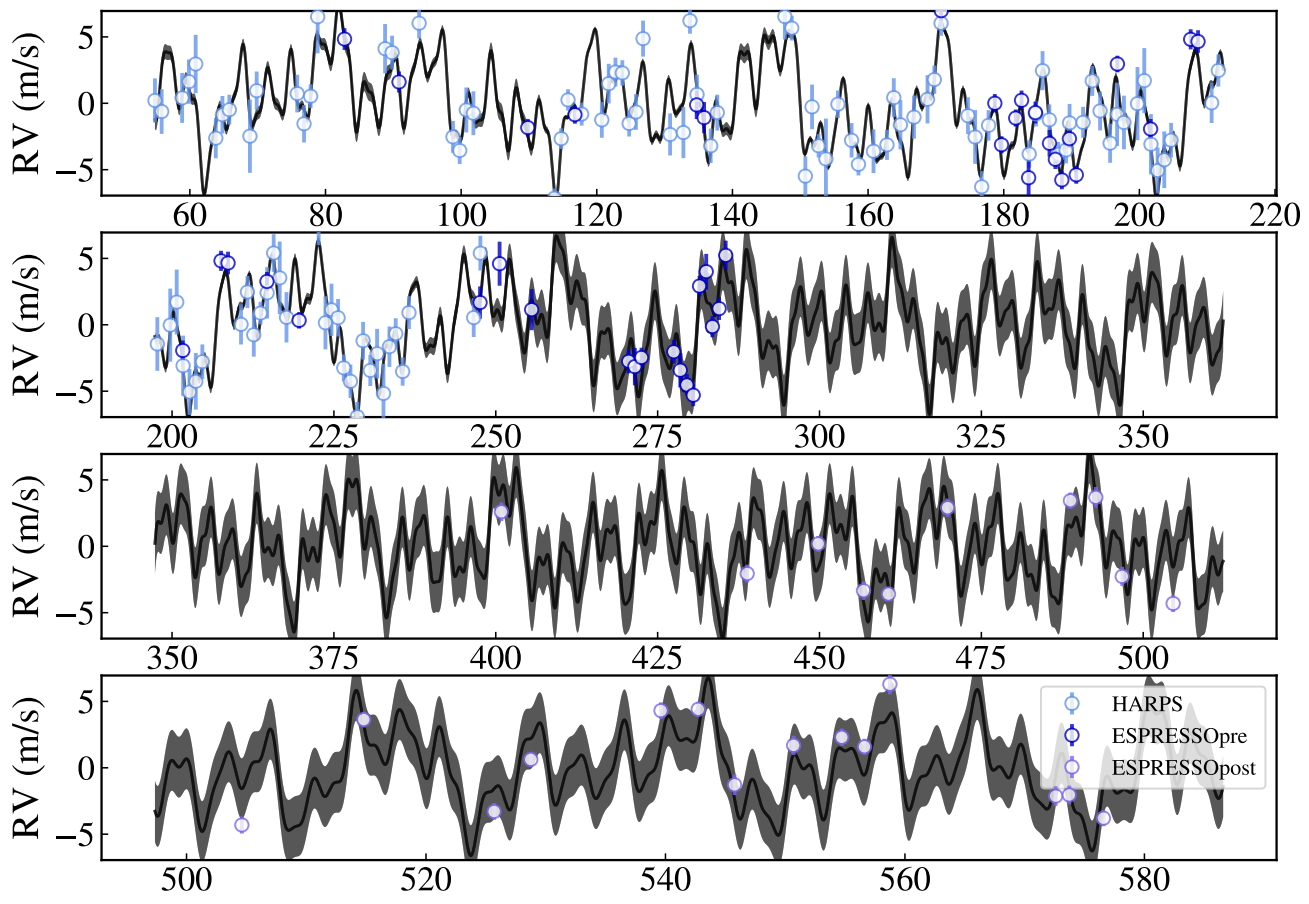

**FIGURE 4** Overview of the full RV data and model including a GP (black line and shaded region) of the best fitting six planet model. The time is in (BJD -2458354). The variance of the model is shown as the shaded region.

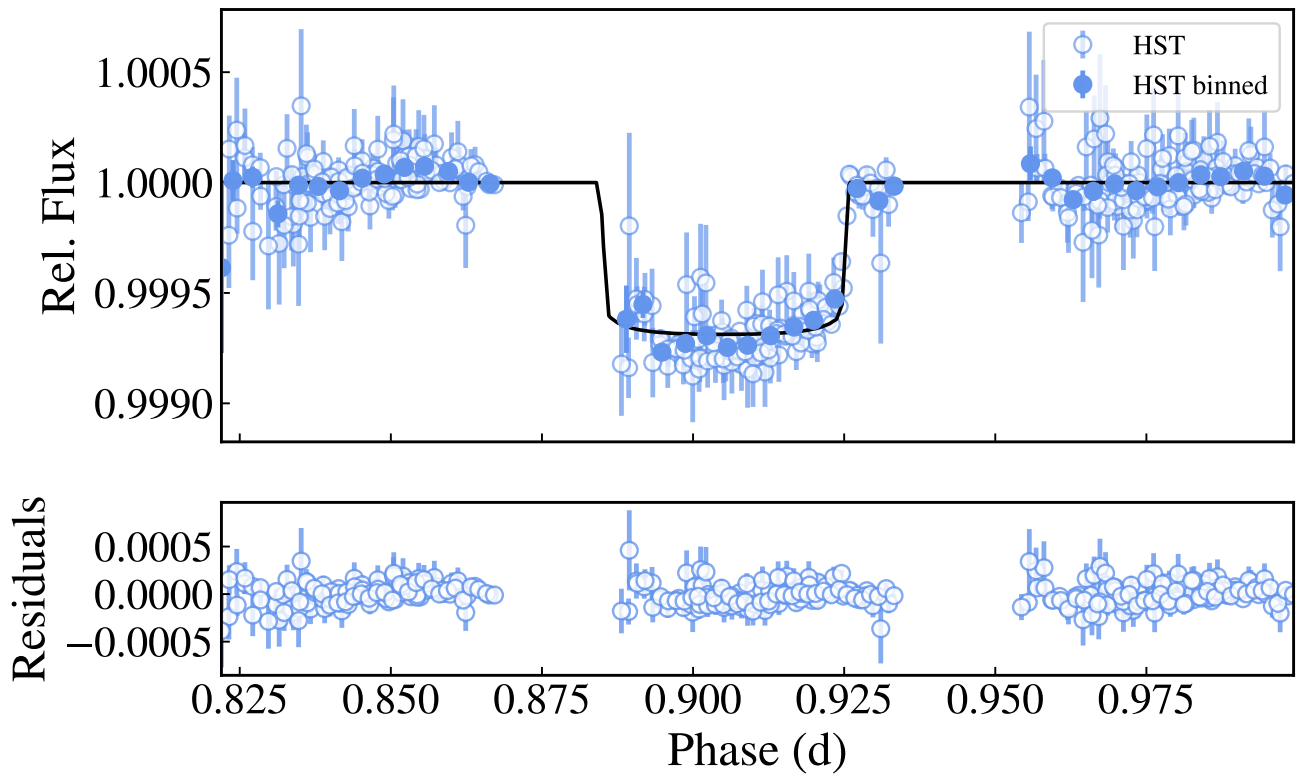

**FIGURE 5** Phase folded transit of L 98-59 b of the binned and unbinned HST data and residuals. The best fitting six planet model is shown by the black curve.

## 5 | DISCUSSION

### 5.1 | Stellar rotation

When comparing the different GP kernels used in the analysis the kernels we find some general trends. For all models using the sSHO kernels a stellar rotation period near $37 \pm 3$ d is preferred. This holds true even when using the very wide prior, with the only exception being the four planet model where the GP tried to include the signature of the fifth planet. For the cdSHO and the dSHO kernels a $P_0$ of $76 \pm 4$ d is preferred, but all of these kernels had their main power in the second harmonic ($f \gtrsim 17$). According to Perger et al. (2021) this could indicate a complex distribution of star spots along two active longitudes, analogous to the sun. Perger et al. (2021) additionally find that the `celerite` SHO kernel tends to prefer a rotation period at the second harmonic of the true rotation period. With the definition of the dSHO and cdSHO we can avoid this aliasing and compare the power of the leading period and the first harmonic. The evidences for the noise models are summarised in Table C1 .

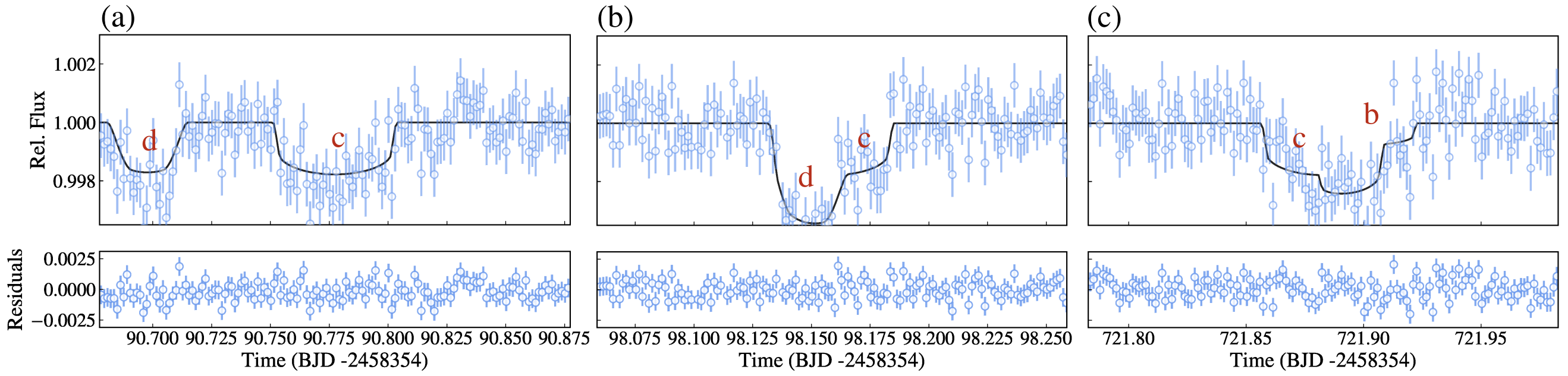

**FIGURE 6** TESS light curves for some selected transits, with the transiting planets marked in red. *(a)*: two distinct transits of L 98-59 c and d. *(b)*: overlapping transit of L 98-59 c and d. *(c)*: overlapping transit of L 98-59 b and c.

Our result for the stellar rotation period was examined using the $\ell$1-periodograms (Hara et al., 2017) – which were set up with a correlated noise model and log-uniform priors. The $\ell$1-periodograms resulted in evidence for either rotation period, where the highest power of the correlated noise was either around 34.7 d or 68.3 d with almost identical cross-validation scores. With the argumentation from Perger et al. (2021), this reinforces our confidence in the rotation period around $76\pm4$ d, but highlights the importance of selecting appropriate kernels when modelling stellar activity and the challenges in determining rotation periods for stars with low activity levels.

H21 used the TESS lightcurves in combination with additional observations of EVRYSCOPE and propose a rotation period of 39.6±2.2 d. However, TESS data – as the authors also mention – is not well suited for such a fit, and only the combination of data yielded a result. In addition to the best fit in our data, the activity indicators in the line indices $H_\alpha$ and the S-Index reinforce our confidence in a rotation period of $76\pm4$ d. We furthermore investigated the stellar activity in the XUV and the relation with the stellar rotation. According to Magaudda et al. (2022) and their published data (Magaudda et al., 2021), L 98-59 has a XUV luminosity of $\log(L_X)[10^{-7}\,\mathrm{W}] = 26.48$. Comparing this to the findings from Magaudda et al. (2020) (figure 5 therein) a rotation period of the star around 80 d is preferred as well.

## 5.2 | Five planet model

As summarised in Table 4 , we find significant evidence for the previously detected fifth planet in the L 98-59 system. The GLS periodograms of the data, the model, the GP, and the residuals of the five planet model are shown in Fig. 2 (*middle* panel). When comparing the GLS periodograms to the ones from the four planet model the peak of the fifth planet was absorbed in the Keplerian orbit model and not in the GP as in the four planet model. The average $\Delta \ln \mathcal{Z}$ increase of 5.7 offers enough evidence to claim that the signature truly has planetary nature. In addition to our analysis with `eff` we examine the $\ell$1 periodograms which were set up using the RV data from HARPS, $\mathrm{ESPRESSO_{pre}}$ and $\mathrm{ESPRESSO_{post}}$. The offsets were individually corrected and added as free parameters in the $\ell$1 routine alongside a correlated noise model as described by the authors of Hara et al. (2017). The $\ell$1-periodogram of the data reduced by the four confirmed planets and the strongest correlated noise shows the highest amplitude coefficient at 23.1 d with a $\log_{10}$FAP of −5.2. Contrary to the findings in D21, our analysis shows low eccentricities for all planets in the system (Table 5 ). This result aligns with the expectations from D21 given the system's multiplicity and inter-planetary interactions.

Fromont et al. (2024) place the optimistic HZ for this system from 0.086 au to 0.224 au, without specifying the Bond albedo explicitly. The conservative HZ in their model ranges from 0.109 au to 0.212 au. This estimation is based on the XUV flux to the planets of this system and places the new planet L 98-59 f close to the inner edge of the conservative habitable zone (HZ) while no other planet would even be in the optimistic HZ. A similar estimation using the spectrum of L 98-59 is also presented in D21.

Estimating the equilibrium temperature for the planets with a simple flux model – assuming an isotropic heat redistribution and a Bond albedo of 0 – lead to a zone of liquid surface water from 0.061 au to 0.114 au, placing planets e and f in the HZ.

The system is scheduled to be observed with the LIFE mission and the HWO. The HWO aims for direct imaging of the planets in this system. This will allow to get estimates on the atmospheres and possibly bio markers of the planets, even the non-transiting ones, which is especially interesting for the outermost planet, since it is located in both estimations of the HZ.

**TABLE 4** Difference in the logarithmic likelihood ($\Delta \log \mathcal{L}$), logarithmic evidence ($\Delta \ln \mathcal{Z}$) and the difference in standard deviations $\sigma$ for the different noise and planet models. The reference for each model comparison is the N-1 planet model.

| Kernel | GP period [d] Prior | GP period [d] Posterior | # | $\Delta \log \mathcal{L}$ | $\Delta \ln \mathcal{Z}$ | $\sigma$ |
|---|---|---|---|---|---|---|
| None | | | 4P | | | |
| | | | 5P | 24.3 | 11.3 | 4.4 |
| | | | 6P | 4.7 | 7.5 | 3.5 |
| sSHO | $\mathcal{U}(25, 500)$ | $28.0^{+2.4}_{-1.7}$ d | 4P | | | |
| | | $35.3^{+3.8}_{-3.2}$ d | 5P | 8.2 | 0.4 | 0.4 |
| | | $35.3^{+3.8}_{-3.2}$ d | 6P | 11.0 | 10.5 | 4.1 |
| sSHO | $\mathcal{N}(39, 5)$ | $30.7^{+3.3}_{-2.5}$ d | 4P | | | |
| | | $37.7^{+3.2}_{-3.2}$ d | 5P | 15.7 | 7.9 | 3.6 |
| | | $37.1^{+3.0}_{-2.7}$ d | 6P | 12.2 | 5.6 | 2.9 |
| sSHO | $\mathcal{N}(78, 5)$ | $77.3^{+4.8}_{-5.0}$ d | 4P | | | |
| | | $76.7^{+4.9}_{-4.9}$ d | 5P | 18.8 | 10.4 | 4.2 |
| | | $77.2^{+4.8}_{-5.0}$ d | 6P | 9.9 | 6.4 | 3.1 |
| cdSHO | $\mathcal{N}(39, 5)$ | $46.6^{+2.6}_{-2.1}$ d | 4P | | | |
| | | $46.2^{+3.4}_{-2.8}$ d | 5P | 5.4 | 7.9 | 3.6 |
| | | $44.7^{+4.0}_{-3.4}$ d | 6P | 14.2 | 8.3 | 3.7 |
| cdSHO | $\mathcal{N}(78, 5)$ | $69.8^{+4.6}_{-4.6}$ d | 4P | | | |
| | | $75.0^{+4.1}_{-4.1}$ d | 5P | 12.4 | 8.7 | 3.8 |
| | | $75.4^{+3.9}_{-3.9}$ d | 6P | 12.6 | 8.6 | 3.7 |
| dSHO | $\mathcal{N}(39, 5)$ | $47.8^{+4.2}_{-4.4}$ d | 4P | | | |
| | | $44.7^{+7.6}_{-3.6}$ d | 5P | 7.3 | 3.3 | 2.1 |
| | | $42.8^{+9.1}_{-2.7}$ d | 6P | 13.5 | 7.0 | 3.3 |
| dSHO | $\mathcal{N}(78, 5)$ | $72.7^{+5.0}_{-5.6}$ d | 4P | | | |
| | | $76.8^{+4.2}_{-4.3}$ d | 5P | 13.3 | 9.0 | 3.8 |
| | | $76.6^{+4.1}_{-4.2}$ d | 6P | 17.2 | 10.7 | 4.2 |

## 5.3 | Six planet model

For the best fitting six-planet model we find a significant evidence increase of $\Delta \ln \mathcal{Z} = 10$ (Table C1 ). The resulting corner-plots can be found in APPENDIX B:. We further validate the robustness of the signal using injection-retrieval tests, and showing, that it is not forced by the choice of priors. This results in the new sixth planet candidate with an orbital period of $1.7361^{+0.0007}_{-0.0008}$ d closest to the host star. Assuming a maximum planetary density comparable to K2-38b (Toledo-Padrón et al., 2020) $11\,\mathrm{g\,cm^{-3}}$ and combining it with the minimum mass $M_c \sin i = 0.57 \pm 0.12\,M_\oplus$ the planetary radius would need to be around $0.66\,R_\oplus$. Such a planet would lead to significant transit signals. However, we are unable to find a transit signature in the light curves and hence expect an inclination offset of $\pm 4.2°$. The fact that the new candidate would be the closest one to the host star and show a larger inclination in comparison to the other planets strengthens the suspicion that this systems dynamics are much more complex than previously thought. However, this is not studied in more depth in this paper.

**TABLE 5** Eccentricities of the best fit models with five and six-planets as well as the results from D21.

| Planet | 5P (D21) | 5P | 6P |
|---|---|---|---|
| 06 | | | $0.03^{+0.04}_{-0.02}$ |
| b | $0.103^{+0.117}_{-0.045}$ | $0.012^{+0.017}_{-0.009}$ | $0.015^{+0.018}_{-0.012}$ |
| c | $0.103^{+0.045}_{-0.058}$ | $0.018^{+0.030}_{-0.014}$ | $0.02^{+0.02}_{-0.02}$ |
| d | $0.074^{+0.057}_{-0.046}$ | $0.008^{+0.013}_{-0.007}$ | $0.013^{+0.016}_{-0.010}$ |
| e | $0.128^{+0.108}_{-0.076}$ | $0.03^{+0.05}_{-0.02}$ | $0.04^{+0.06}_{-0.03}$ |
| f | $0.21^{+0.17}_{-0.11}$ | $0.06^{+0.08}_{-0.05}$ | $0.07^{+0.08}_{-0.06}$ |

As discussed in Sec. 5.1 we find evidence for a rotation period of the host star around $76 \pm 4$ d. As summarised in Table 4 the choice of kernel and priors did not significantly effect the evidence for the sixth planet candidate. For all tested noise models the sixth planet candidate was at least $\Delta \ln \mathcal{Z} = 5.6$ better than the five planet model, which is significant (according to Trotta (2008)).

When studying the $\ell 1$ periodogram of the system with masked periods for the known five planets, we find strong peaks near the presumed rotation period as well as near half a year. Nonetheless, the period with the lowest FAP corresponds to the period of the sixth planetary candidate. Additionally, there is a peak of the same strength present at the alias period of the sixth planet candidate (2.34 d) which has a FAP of 1. This again supports the claim that there the signature at 1.736 d is not only an artefact from the observation, data reduction or the preceded orbit modelling.

As for the five planet best fit model, the eccentricities of the six-planet model are all low (see Table 5 ) and in agreement with the ones of the five planet model. Generally, the orbital and planetary parameters of the present results are in agreement with previous studies. The error on the orbital periods of the confirmed planets could be reduced by up to an order of magnitude, and at least to the same degree. This re-analysis is a great step to better understand the evolution and dynamics in planetary systems with large multiplicities such as the L 98-59 and TRAPPIST-1 system.

As highlighted by Rajpaul, Barragán, and Zicher (2024), a simple fit resulting in a non-zero radial velocity is insufficient evidence to claim a confirmation of an additional planet. When searching for the new candidate with wide period priors ($\mathcal{U}(1.1\,\mathrm{d}, 1.9\,\mathrm{d})$), the period of $1.7362^{+0.0006}_{-0.0012}$ d can be found but the evidence increase is insignificant. Nonetheless, this new candidate can be seen as a highly significant fit since the log-likelihood increase is significant and independent of the choice of prior distribution.

Turtelboom, Dietrich, Dressing, and Harada (2024) examined the L 98-59 system with empirical models, trying to predict additional planets in known TESS multi-planet systems. In their evaluations using the period ratio models (PRM, Fig. 5 in Turtelboom et al., 2024) the posterior probability distributions from `DYNAMITE` has the highest two peaks at $\sim 5$ d and around 1.6 d. We do not find any indication of the $\sim 5$ d signal. Our signal at 1.74 d could explain the peak in the PRM around 1.6 d. However, further investigations are necessary to confirm or reject this signal.

## 5.4 | Stability with a sixth planet

For a realistic estimation of the orbital and planetary parameters we use a stability analysis with the python package `SPOCK` Tamayo et al. (2020). `SPOCK` (Stability of Planetary Orbital Configurations Klassifier) is a package build for '*predicting the long-term stability of compact multi-planet systems*' (Tamayo et al., 2020, Title). It utilises machine learning models to estimate the stability of planetary systems after $10^9$ planetary orbits.

We use `SPOCK` on our posterior samples and estimated the stability of the orbits to get a more realistic approximation of the orbital parameters. We could not find constraints from the stability analysis in the posteriors of the model with the best Bayesian evidence.

Comparing L 98-59 with the well-studied TRAPPIST-1 system we see similarities but also significant differences. The orbital separations in units of the mutual Hill radii ($\Delta(R_H)$) of the planets in the system show differences, as the lowest separation in the L 98-59 system is $\Delta(R_H) = 11.9$ (planet candidate 06 and planet b), while the separations in TRAPPIST-1 range from $\Delta(R_H) = 6.6$ to $\Delta(R_H) = 12.8$. This implies that there are only minimal interactions between the planets in this system. The new planetary candidate L 98-59.06 increases the tension in the system, since the next closest separations in units of the mutual Hill radii are between the planets d and e ($\Delta(R_H) = 14.1$), and e and f ($\Delta(R_H) = 14.3$).

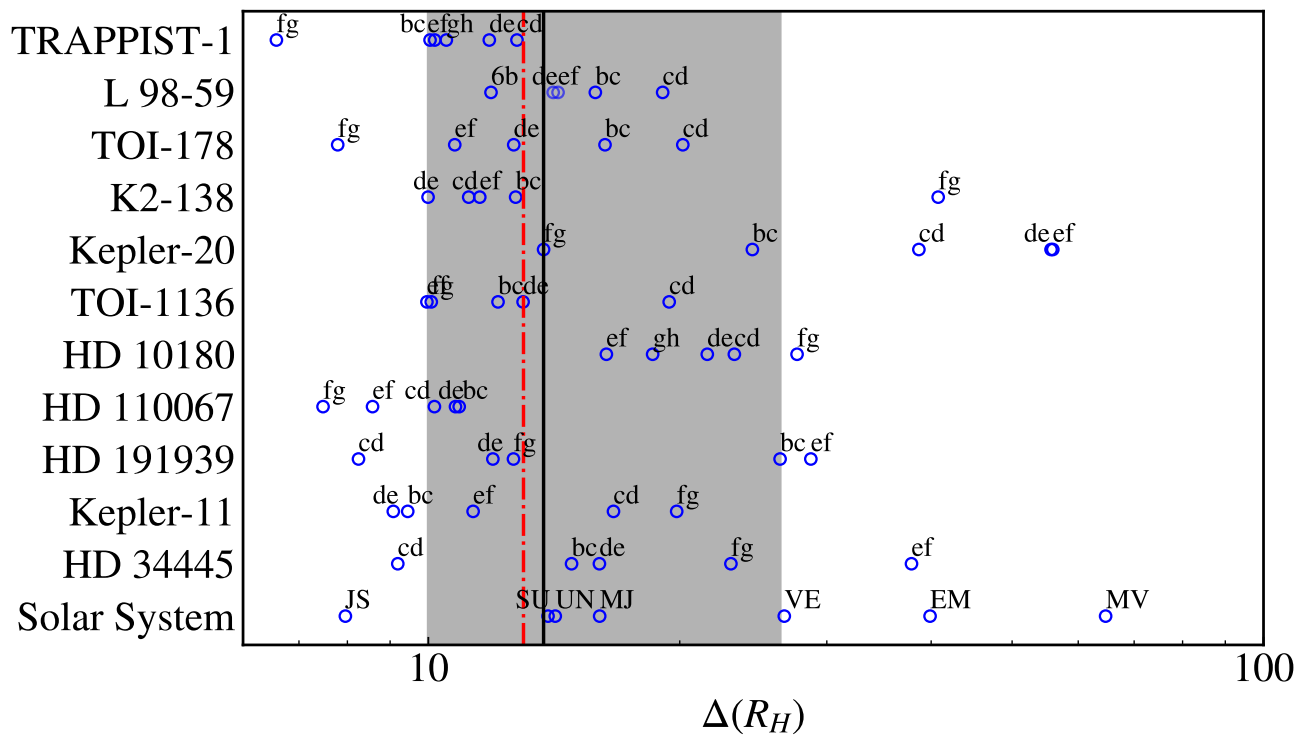

**FIGURE 7** Separations in units of the mutual Hill radius of all the planetary systems with six or more known planets. The median of the separation for the multi-planet systems (solid black line and grey shaded area for the 16th-84th percentile) is centred close to the stability limit of $\Delta(R_H) = 13$ (dash-dotted red line) from Dreizler et al. (e.g. 2024); Weiss et al. (e.g. 2018).

Figure 7 shows the orbital Separation in mutual Hill radii of the planetary systems with 6 and more planets, based on Figure 6 in Dreizler et al. (2024). The labels next to the open circles indicate the planet pair for this separation. The black, solid line corresponds to the median ($\Delta(R_H) = 14$) of the separation, the shaded region to the 16% − 84% quantiles of the separations which are $\Delta(R_H) = 10$ and $\Delta(R_H) = 28$ respectively. The red dashed-dotted line at $\Delta(R_H) = 13$ corresponds to the limit where the orbit crossing time scale is on average above $10^9$ orbits (Chambers, Wetherill, & Boss, 1996; Gratia & Lissauer, 2021; Ormel, Liu, & Schoonenberg, 2017; Rice & Steffen, 2023; Smith & Lissauer, 2009).

## 5.5 | System architecture and composition

The planetary system of L 98-59 is a tightly packed, close-in system with some planets near orbital period commensurabilities. The new sixth planet candidate falls in line with its companions. The new planet candidate L 98-59.06 and L 98-59 c are near a 2:1 (2.1:1), L 98-59 c and d are near a 2:1 (2.01:1), and L 98-59 d and f near a 3:1 (3.1:1) orbital period commensurability.

L 98-59.06 and planet b have relatively low masses of $0.57 \pm 0.12\,M_\oplus$ sin i and $0.33 \pm 0.12\,M_\oplus$, respectively. Nonetheless, their posteriors exclude values of zero with $3\sigma$ confidence (L 98-59.06) and $2\sigma$ confidence (L 98-59 b), respectively. The other four planets have masses ranging from $1.9 \pm 0.2\,M_\oplus$ (L 98-89 d) to $3.0 \pm 0.5\,M_\oplus$ (L 98-59 f), which is consistent with the 'peas-in-a-pod' theory from W18. For multi-planet systems with three planets or more, W18 predict a ratio of period ratios $\mathcal{P} = (P_{j+2}/P_{j+1})/(P_{j+1}/P_j)$ of $1.03 \pm 0.27$, and the planets in the L 98-59 system nicely fall in line with this prediction. In addition to $\mathcal{P}$, Table 6 lists the rate of radii of the three transiting planets, the separations in mutual Hill radii (for a visualisation of the separations and overlap see Fig. 7 ), and the difference in equilibrium temperature of neighbouring planets $\Delta T_{\mathrm{eq}}$, with comparison values from D21 and the references from W18. As discussed in D21 we find that the $\Delta T_{\mathrm{eq}}$ scales with the ratios in planetary radii, as predicted by W18.

**TABLE 6** Peas-in-a-pod metrics analogous to Table 2 in D21.

| Metric | this work | D21 | W18 distribution |
|---|---|---|---|
| $R_c/R_b$ | 1.55 | 1.669 | $1.14 \pm 0.63$ |
| $R_d/R_c$ | 1.16 | 1.077 | |
| $(P_c/P_b)/(P_b/P_{06})$ | 1.26 | | $1.03 \pm 0.27$ |
| $(P_d/P_c)/(P_c/P_b)$ | 1.232 | 1.232 | |
| $(P_e/P_d)/(P_d/P_c)$ | 0.855 | 0.851 | |
| $(P_f/P_e)/(P_e/P_d)$ | 1.04 | 1.053 | |
| $\Delta(R_H)_{06,\,\mathrm{b}}$ | 11.89 | | Maximum at 20 and tail to large separations. Stable for $> 10^9$ orbits above 13 |
| $\Delta(R_H)_{b,c}$ | 15.85 | 15.260 | |
| $\Delta(R_H)_{c,d}$ | 19.18 | 18.414 | |
| $\Delta(R_H)_{d,e}$ | 14.11 | 13.569 | |
| $\Delta(R_H)_{e,f}$ | 14.29 | 14.389 | |
| $\Delta T_{\mathrm{eq}}(06,b)$ [K] | 55 | | Correlated to the radius ratio. |
| $\Delta T_{\mathrm{eq}}(b,c)$ [K] | 94 | 49 | |
| $\Delta T_{\mathrm{eq}}(c,d)$ [K] | 110 | 152 | |
| $\Delta T_{\mathrm{eq}}(d,e)$ [K] | 68 | 119 | |
| $\Delta T_{\mathrm{eq}}(e,f)$ [K] | 61 | 57 | |

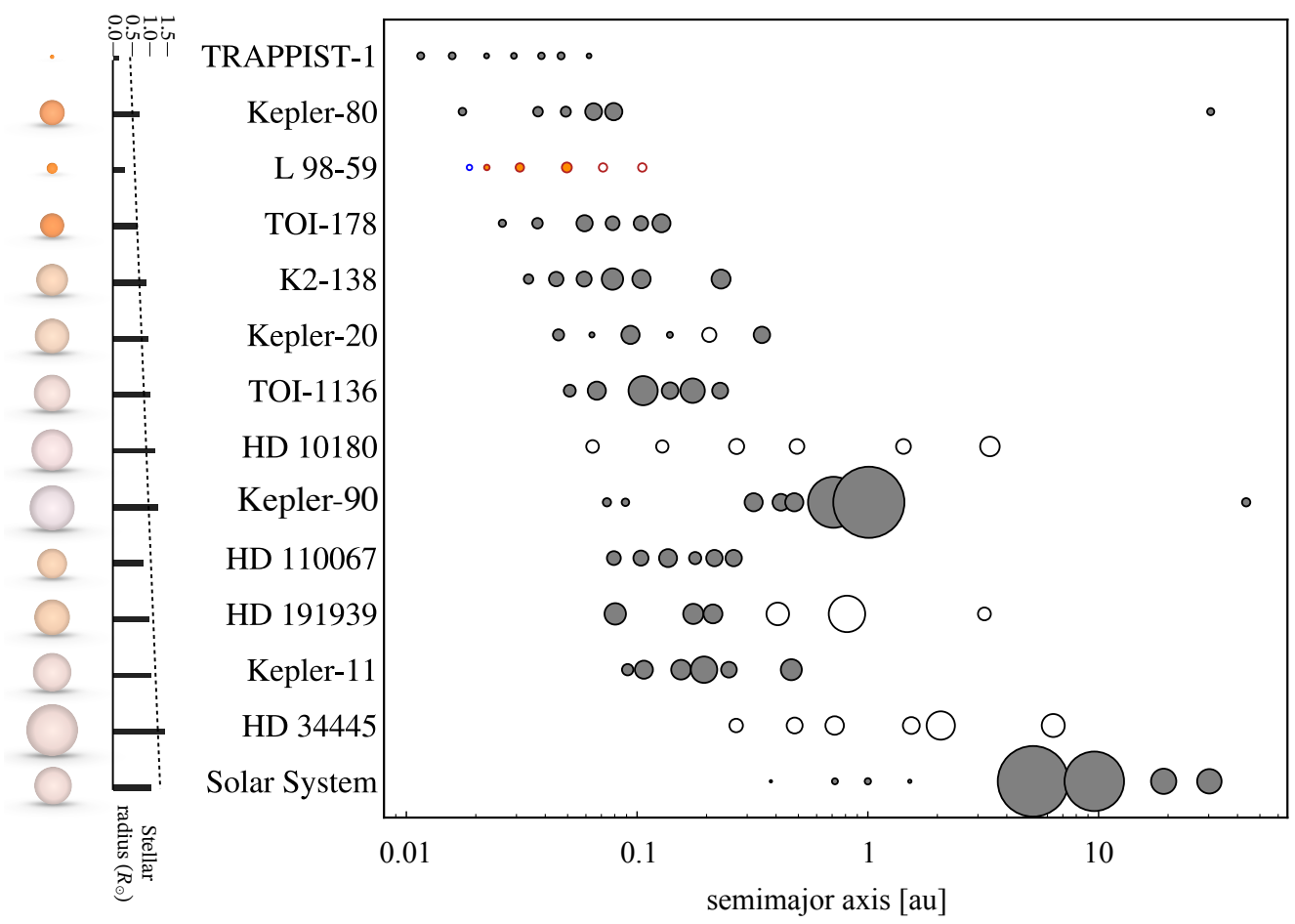

**FIGURE 8** Peas in a pod comparison of all known systems with six or more planets. In the main diagram, grey symbols correspond to the planetary radii, with star-planet distances along the abscissa. The planet candidate is shown in blue, the rest of the L 98-59 system in red. Colours for the stellar symbols in the left column are from Harre and Heller (2021). Stellar radii are to scale. Systems are sorted by increasing semimajor axis of the innermost planet. The vertical histogram shows the stellar radii, and the dashed line illustrates a linear fit.

In Fig. 8 we show a comparison of the planetary architectures of 12 planetary systems with at least six confirmed planets at the time of writing[5]. This collection is expanded by the solar system and the L 98-59 system. They are sorted by increasing semimajor axis of the innermost planet. Planetary radii are symbolised by the size of the grey filled planetary circles. The planets where no planetary radius is known are shown as empty circles with radii approximations from their mass (assuming terrestrial composition). The planetary radii and masses in the L 98-59 system are highlighted in red, and the new candidate L 98-59.06 is indicated in blue. The visual appearance of the host stars are according to the digital colour codes of stars from Harre and Heller (2021), Table 5 therein. Stellar radii are to scale, but not in any relationship with the planetary symbols. Although the systems are sorted according to a planetary characteristic, the orbital semimajor axis of the innermost planet, we notice a faint trend from late-type stars to early-type stars and increasing stellar radius from top to bottom. This trend is highlighted in the vertical histogram in Fig. 8 , in which we plot the stellar radius for each system together with a linear fit (dashed line).

Figure 9 depicts a mass-density plot and is a reproduction from Luque and Pallé (2022) and contains data points from planets orbiting M dwarfs. The densities are scaled with the model for an earth-like composition. It includes planets with terrestrial composition (brown) and water worlds (blue), as well as predictions for an earth-like density (solid black line) and predictions for a rocky composition with a water content of 50% (dashed black line). Marked in black solid circles are the planets of the TRAPPIST-1 system, as well as L 98-59 b, c and d. The non-transiting planets of this system are indicated by the shaded and hashed areas. As shown in Fig. 9 the three transiting planets follow the predictions from Luque and Pallé (2022) for compositions of planets around M dwarfs;

[5] Query from `https://exoplanetarchive.ipac.caltech.edu/cgi-bin/TblView/nph-tblView?app=ExoTblsŹconfig=PS` 12/2024

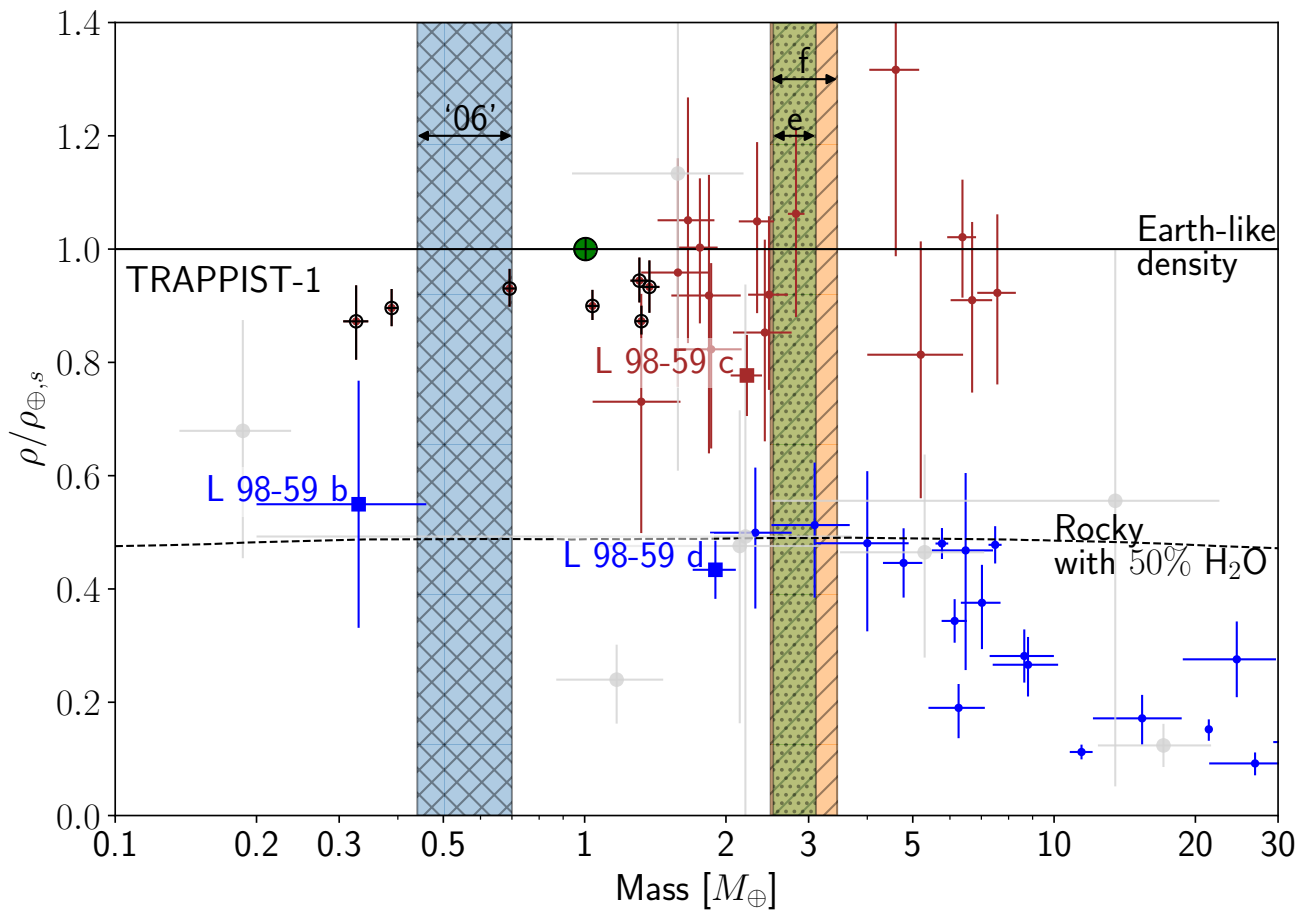

**FIGURE 9** Mass-Density plot for planets orbiting M dwarfs, reproduced from Luque and Pallé (2022) with updated values for (L 98-59 b, c and d, squares), as well as the mass regions for the non-transiting planets L 98-59 e (dotted, green), f (hatched, orange) and L 98-59.06 (cross-hatched, blue). The data points shown in brown correspond to planets with earth-like densities. The blue data points show planets with rocky cores and 50% water content. The grey data points correspond to planets with mass uncertainties larger than 25%.

L 98-59 c is in the regime of terrestrial planets with a density of $4.9^{+0.6}_{-0.5}$ g cm$^{-3}$. L 98-59 d with a density of $2.7^{+0.4}_{-0.3}$ g cm$^{-3}$ on the other hand is more likely to have a water content closer to 50% alongside the earth-like composition. The third body with constraints on both mass and radius L 98-59 b shows a density of $2.6 \pm 1.0$ g cm$^{-3}$ which is located closer to the densities of a water content around 50% as well. When comparing this to the other planets in this regime it becomes apparent that this combination of low mass and density is uncommon for planets around M-Stars (grey data points in Fig. 9 ). The new planet candidate is located at a mass of $0.58 \pm 0.12\,M_\oplus$ (blue cross-hatched region in Fig. 9 ) and would be one of the lightest planets found with RVs. When following the general trend of small planets around M dwarfs this would imply an earth-like composition. However, this is not necessarily the case, since L 98-59 b diverts from this expectation as well. If L 98-59.06 would follow the predictions for an earth-like composition it would have an approximate radius range of $0.8\,R_\oplus$ to $0.9\,R_\oplus$.

## 6 | CONCLUSION

Here we confirm the presence of the previously claimed RV signal (D21) of a fifth planet in the L 98-59 system (planet f) with an orbital period of $23.07 \pm 0.09$ d and a minimum mass of $3.0 \pm 0.5\,M_\oplus$. We use a different extraction method for the HARPS data (SERVAL) and new observations from 12 TESS sectors of the transits of planets b, c, and d plus five previously observed HST transit observations of planet b. Our statistical analysis yields significant evidence ($3.3 \leq \Delta \ln \mathcal{Z} \leq 11.3$, $2.1 \leq \sigma \leq 4.4$, depending on GP parameterization) for the fifth planet.

We examine various indicators of stellar activity to exclude a stellar origin of the signal. Our investigation of the stellar activity indicators and of the GP modelling yields moderate evidence for a stellar rotation period around $76 \pm 4$ d, in agreement with previous findings. The fact that the main power of the dSHO kernels remained mostly in the second harmonic indicates a complex star spot pattern on the stellar surface (Perger et al., 2021).

We also find an additional signal in the GLS periodograms of the five planet model, which we attribute to a new sixth planet candidate (L 98-59.06). The candidates orbital period is $1.7361^{+0.0007}_{-0.0008}$ d and the RV data suggest a minimum mass of $0.58 \pm 0.12\,M_\oplus$. Estimating the planet radius using the highest known planet density, we expect $R_{06} \geq 0.66\,R_\oplus$. If this planet would be transiting, a signal should be prominently visible. We could not find any evidence for transits of this candidate in the TESS lightcurves and conclude that its inclination has to be misaligned by $\pm 4.2°$ so that no transits are visible.

As a next step towards a consistent characterisation of the system, a fully dynamic study of the planet-planet interaction is necessary. This would allow independent constraints of the planetary masses, better estimates of the planetary radii, and consistent computations of the orbital elements. The L 98-59 system remains a very interesting benchmark for the study of orbital dynamics and for studies of planetary composition and evolution.

## ACKNOWLEDGEMENTS

Funding for the TESS mission is provided by NASA's Science Mission directorate. This research has made use of the Exoplanet Follow-up Observation Program (ExoFOP; DOI: 10.26134/ExoFOP5) website, which is operated by the California Institute of Technology, under contract with the National Aeronautics and Space Administration under the Exoplanet Exploration Program. This paper includes data collected by the TESS mission, which are publicly available from the Mikulski Archive for Space Telescopes (MAST). Based on data obtained from the ESO Science Archive Facility with DOI: `https://doi.eso.org/10.18727/archive/33`. Based on observations collected at the European Southern Observatory under ESO programs 1102.C-0744, 1102.C-0958, 1104.C-0350, 198.C-0838(A), 1102.C-0339(A) and 0102.C-0525.

This research has made use of the VizieR catalogue access tool, CDS, Strasbourg, France.

S. D. acknowledges support from the German Science Foundation (DFG) from grant DR 281/32-1 and DR 281/37-1. R. H. acknowledges support from the German Aerospace Agency (Deutsches Zentrum für Luft- und Raumfahrt) under PLATO Data Center grant 50OO1501.

The authors want to thank the authors of Zhou et al. (2022) who provided the white-lightcurve of the HST transits used in this paper. The authors thank Dr. R. Luque for helpful remarks concerning the reproduction of Fig. 9 .

# APPENDIX A: DETRENDING

We compare the r-spline detrending of the TESS light curves with a median and biweight filtering. We use a window_length of 2 d (all detrending methods), an edge_cutoff of 0 (for median and biweight) and a cval of 5 for the biweight filter. We find that the r-spline has the least influence on the transit shapes and depth while the others show a minimal trend during overlapping transits. Additionally, the r-spline showed a better fit for the trends after breaks in the light curves.

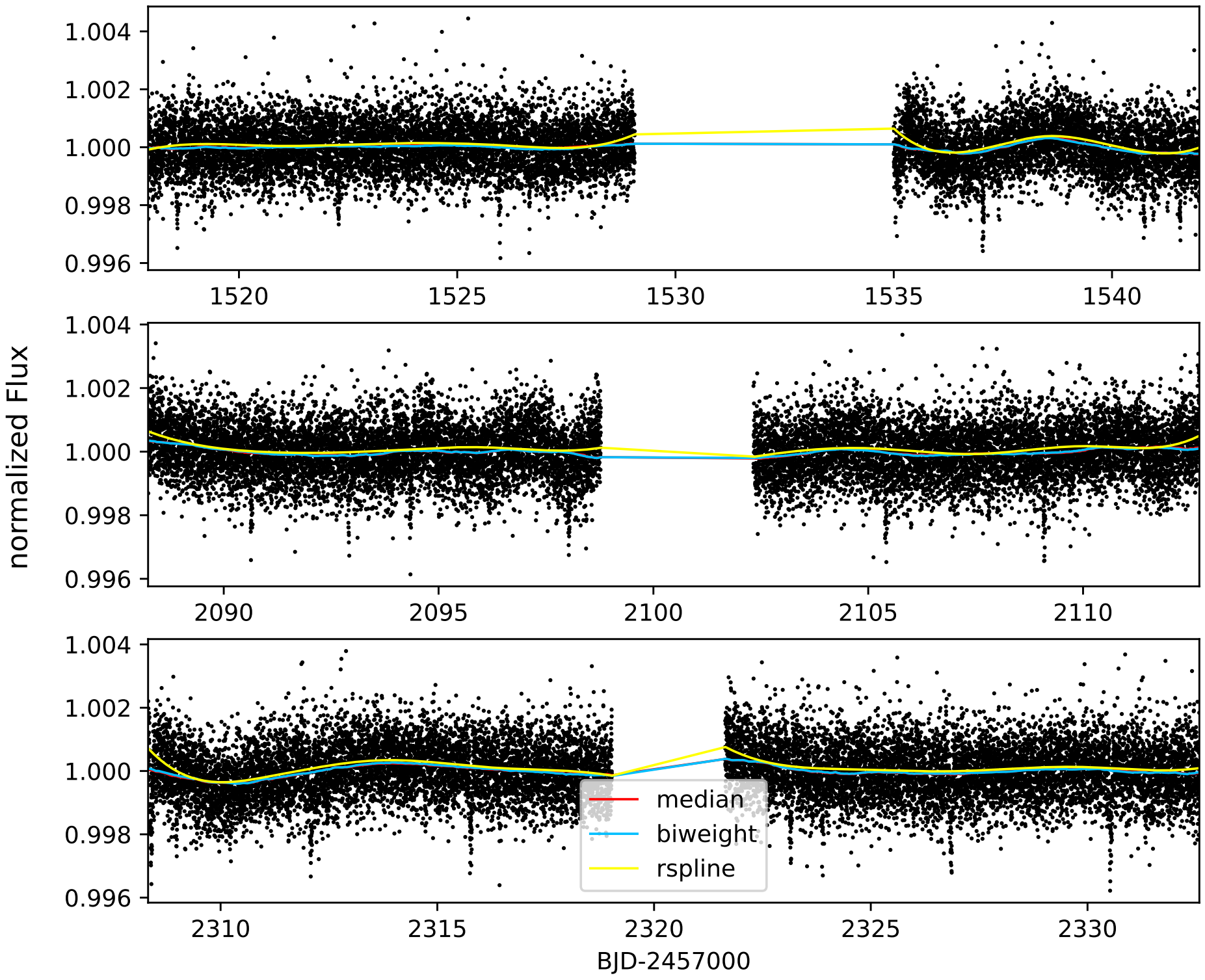

**FIGURE A1** Comparison of the median, biweight and rspline detrending of some selected TESS sectors. *Top* sector 8: with a large gap and some incline at the beginning and end of the gap. Here the rspline best followed the trend, while the other two had some trouble reproducing it. *Middle* sector 29: smaller gap and reasonably constant, but here the different detrending methods showed different trends. The biweight and median seemed to be influenced by the transits and decreased in intensity at these times (between (2105 to 2110) BJD – 2457000). *Bottom* sector 37: generally good and consistent detrending with negligible deviations from one another.

# APPENDIX B: CORNER PLOTS

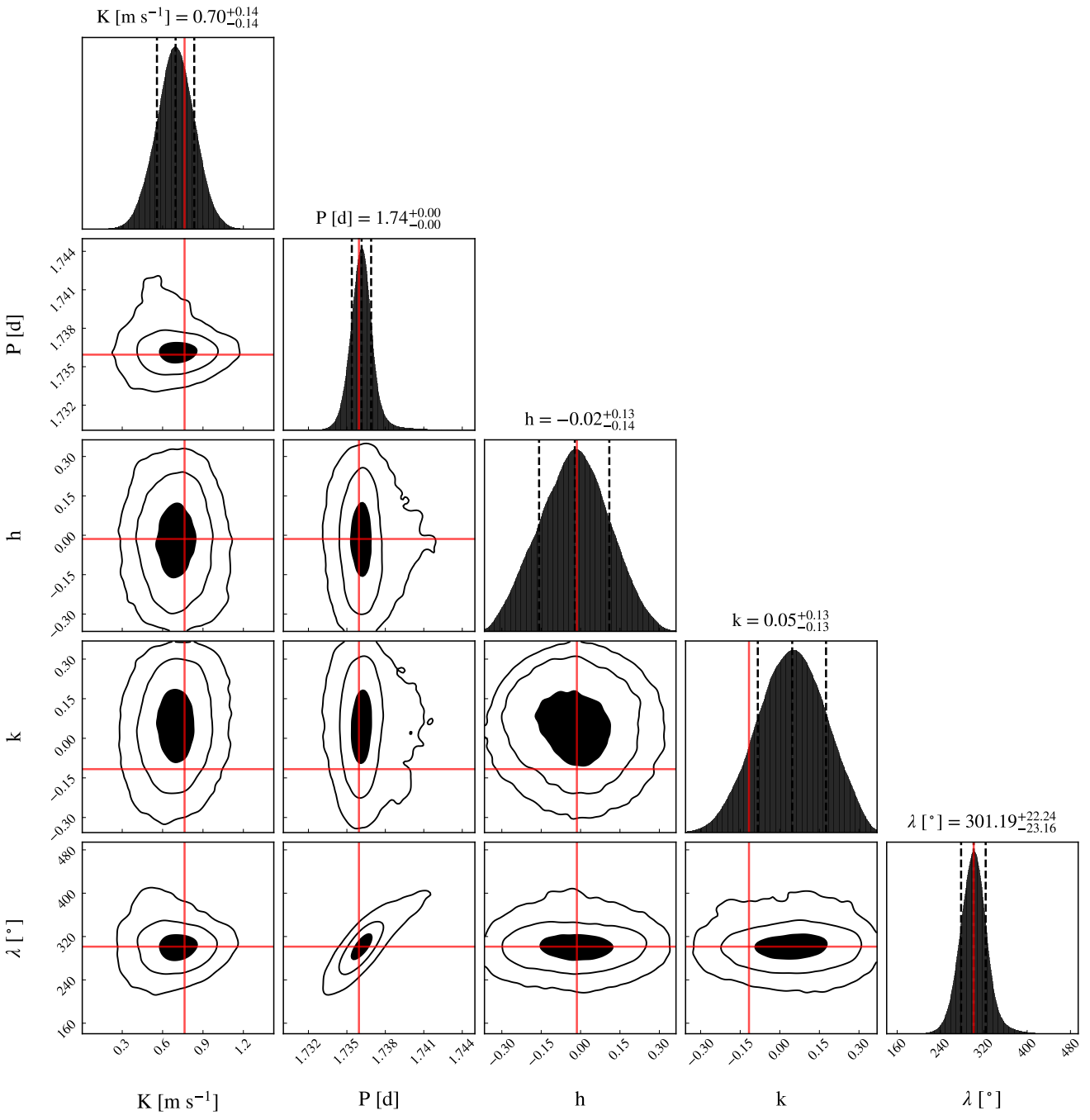

**FIGURE B1** L 98-59.06

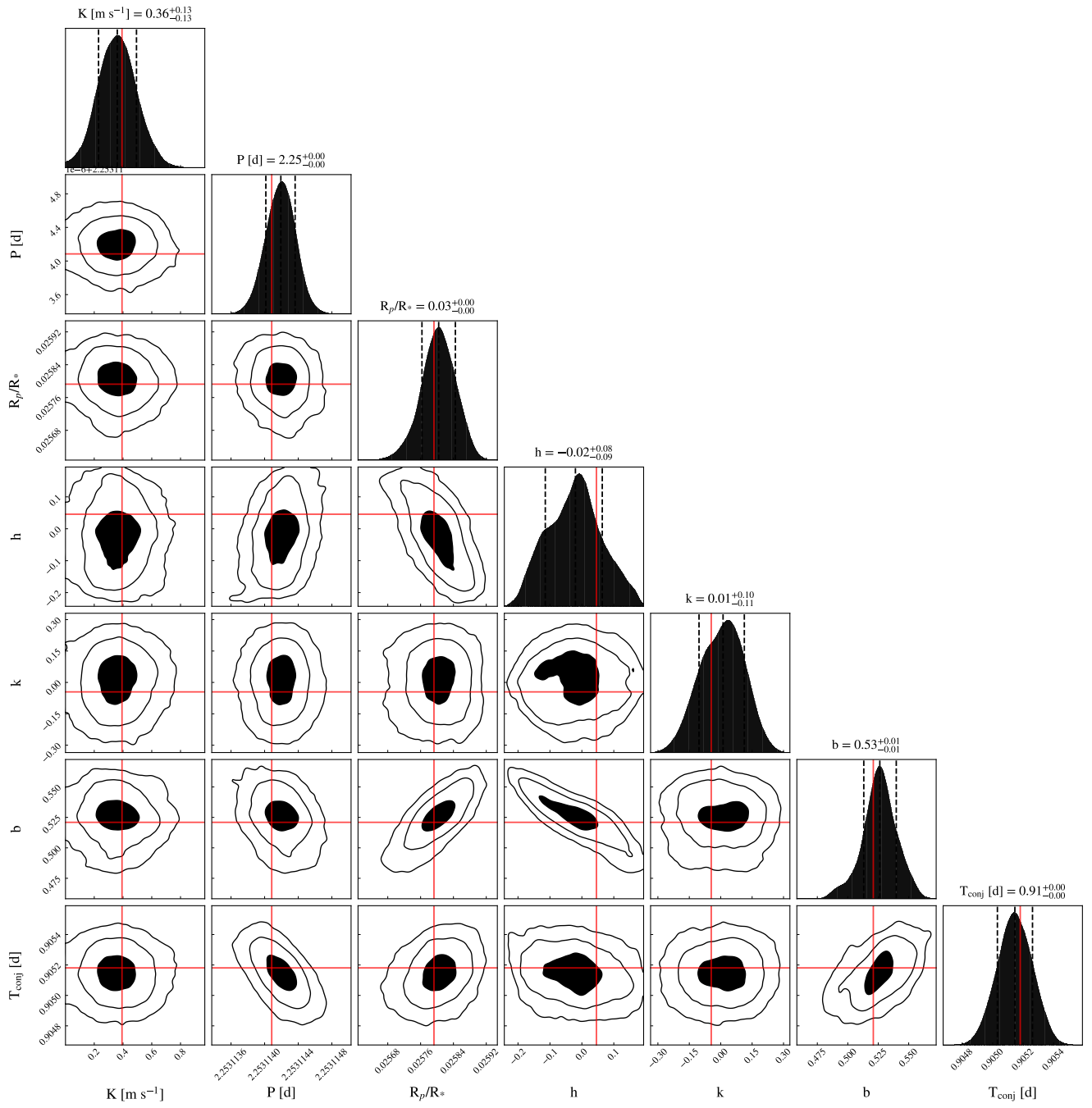

**FIGURE B2** L 98-59 b

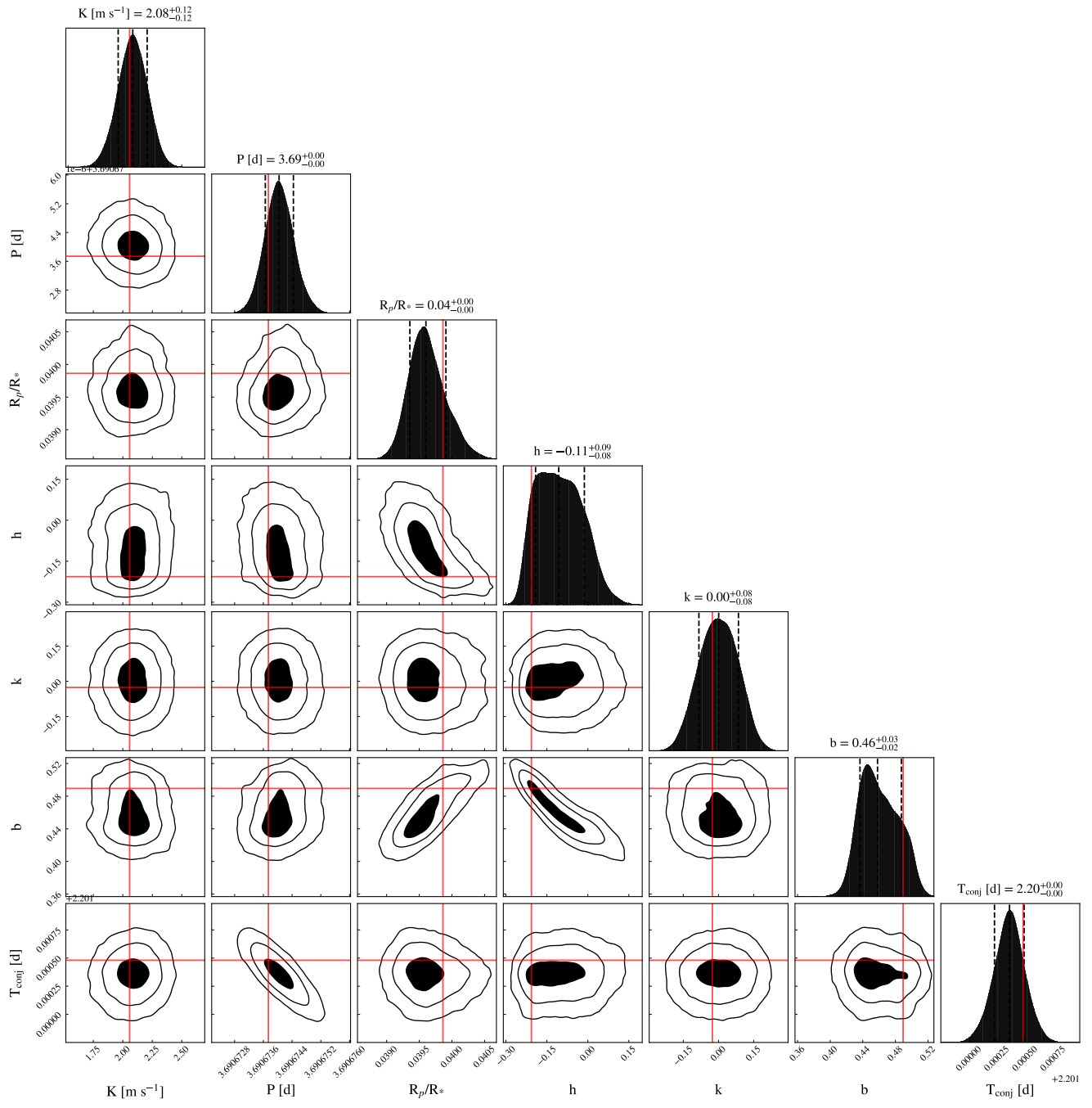

**FIGURE B3** L 98-59 c

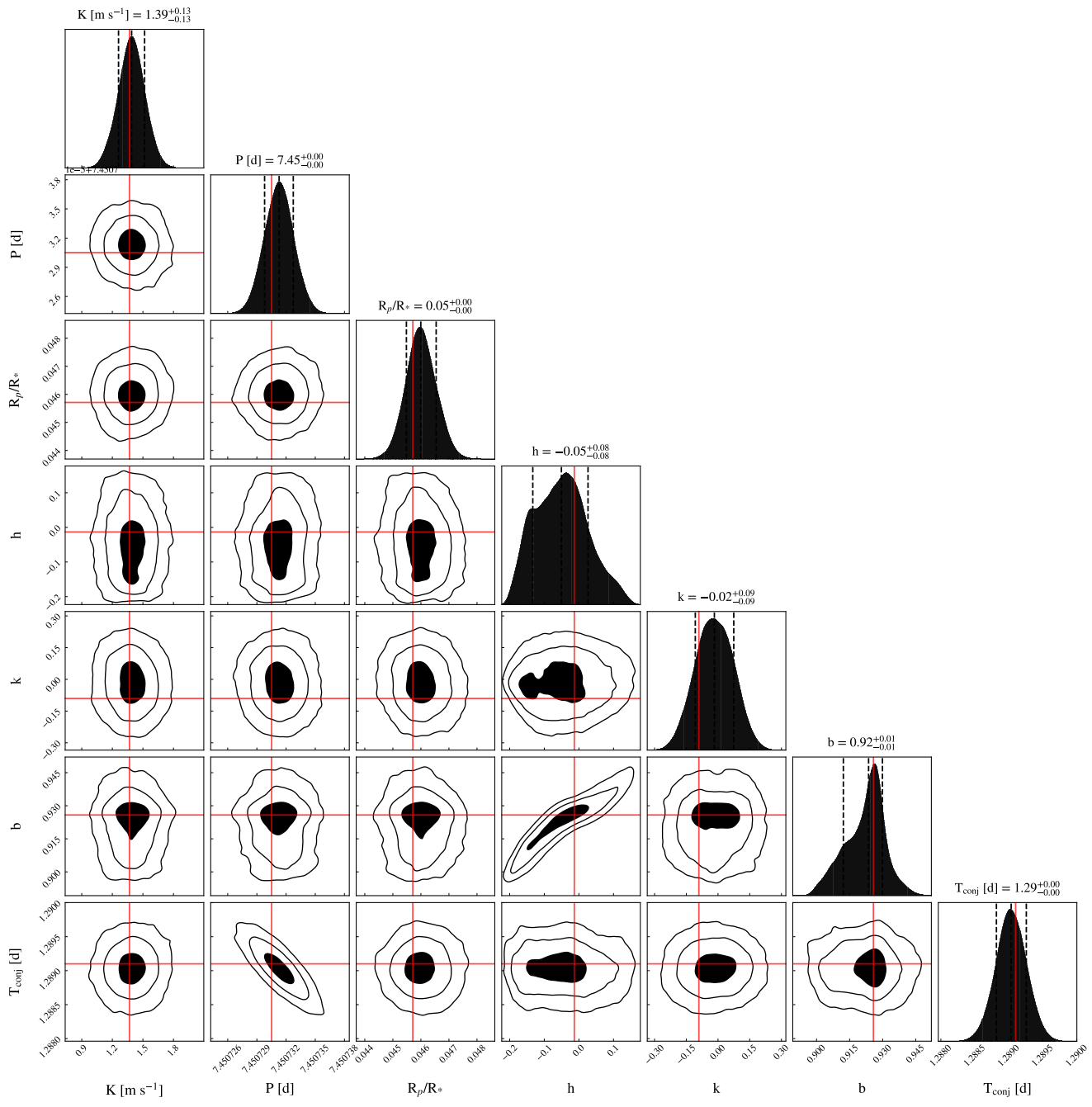

**FIGURE B4** L 98-59 d

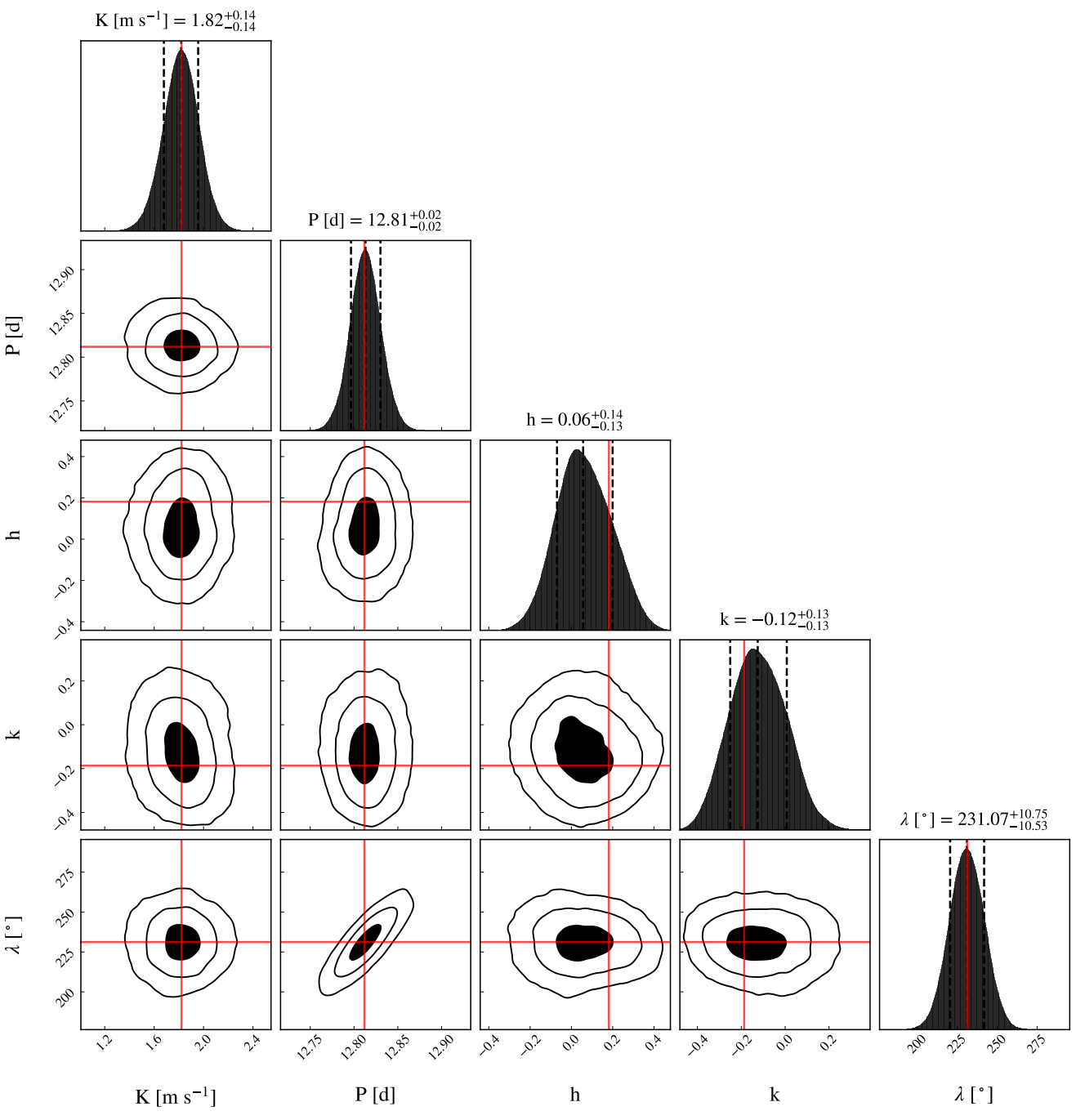

**FIGURE B5** L 98-59 e

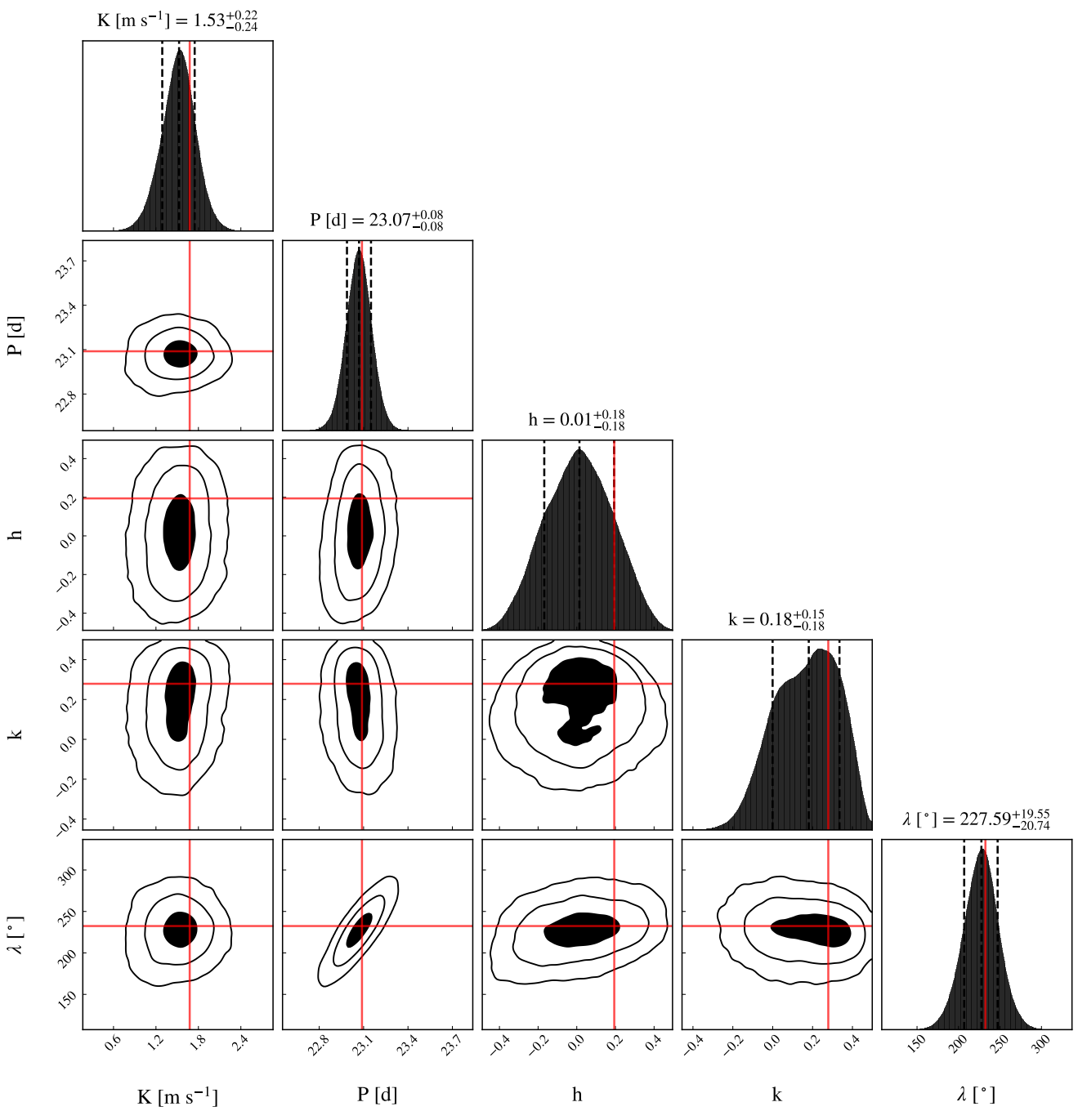

**FIGURE B6** L 98-59 f

# APPENDIX C: TABLES

**TABLE C1** Overview of the difference in logarithmic evidence ($\Delta \ln \mathcal{Z}$) and the logarithmic likelihood ($\Delta \log \mathcal{L}$) for the different GP kernel functions used. The reference is the best 5 Planet model (sSHO, $\mathcal{N}(39,5)$).

| | kernel | GP period Prior | GP period Posterior | $\ln \mathcal{Z}$ | $\Delta \ln \mathcal{Z}$ | $\log \mathcal{L}$ | $\Delta \log \mathcal{L}$ |
|---|---|---|---|---|---|---|---|
| b, c, d, e | None | | | 93.0 | −44.2 | 994.0 | −58.0 |
| | sSHO | $\mathcal{U}(25,500)$ | $28.0^{+2.4}_{-1.7}$ d | 131.5 | −5.7 | 1038.5 | −13.5 |
| | sSHO | $\mathcal{N}(39,5)$ | $30.7^{+3.3}_{-2.5}$ d | 129.3 | −7.9 | 1036.3 | −15.7 |
| | sSHO | $\mathcal{N}(78,5)$ | $77.3^{+4.8}_{-5.0}$ d | 119.6 | −17.6 | 1027.6 | −24.3 |
| | cdSHO | $\mathcal{N}(39,5)$ | $46.6^{+2.6}_{-2.1}$ d | 123.5 | −13.7 | 1040.3 | −11.6 |
| | cdSHO | $\mathcal{N}(78,5)$ | $69.8^{+4.6}_{-4.6}$ d | 125.7 | −11.5 | 1035.0 | −17.0 |
| | dSHO | $\mathcal{N}(39,5)$ | $47.8^{+4.2}_{-4.4}$ d | 126.4 | −10.8 | 1036.1 | −15.9 |
| | dSHO | $\mathcal{N}(78,5)$ | $72.7^{+5.0}_{-5.6}$ d | 127.5 | −9.7 | 1034.1 | −17.8 |
| b, c, d, e, f | None | | | 104.3 | −27.5 | 1018.3 | −33.7 |
| | sSHO | $\mathcal{U}(25,500)$ | $35.3^{+3.8}_{-3.2}$ d | 131.9 | −5.3 | 1046.7 | −5.2 |
| | sSHO | $\mathcal{N}(39,5)$ | $37.7^{+3.2}_{-3.2}$ d | 137.2 | 0 | 1052.0 | 0 |
| | sSHO | $\mathcal{N}(78,5)$ | $76.7^{+4.9}_{-4.9}$ d | 130.0 | −7.2 | 1046.4 | −5.6 |
| | cdSHO | $\mathcal{N}(39,5)$ | $46.2^{+3.4}_{-2.8}$ d | 131.4 | −5.8 | 1045.7 | −6.3 |
| | cdSHO | $\mathcal{N}(78,5)$ | $75.0^{+4.1}_{-4.1}$ d | 134.4 | −2.8 | 1047.4 | −4.6 |
| | dSHO | $\mathcal{N}(39,5)$ | $44.7^{+7.6}_{-3.6}$ d | 129.7 | −7.5 | 1043.4 | −8.5 |
| | dSHO | $\mathcal{N}(78,5)$ | $76.8^{+4.2}_{-4.3}$ d | 136.5 | −0.7 | 1047.4 | −4.6 |
| b, c, d, e, f, 06 | None | | | 111.8 | −25.4 | 1023.0 | −28.9 |
| | sSHO | $\mathcal{U}(25,500)$ | $35.3^{+3.8}_{-3.2}$ d | 142.0 | +4.8 | 1059.5 | +7.5 |
| | sSHO | $\mathcal{N}(39,5)$ | $37.1^{+3.0}_{-2.7}$ d | 142.8 | +5.6 | 1064.2 | +12.2 |
| | sSHO | $\mathcal{N}(78,5)$ | $77.2^{+4.8}_{-5.0}$ d | 136.4 | −0.8 | 1056.3 | +4.3 |
| | cdSHO | $\mathcal{N}(39,5)$ | $44.7^{+4.0}_{-3.4}$ d | 139.7 | +2.5 | 1059.9 | +8.0 |
| | cdSHO | $\mathcal{N}(78,5)$ | $75.4^{+3.9}_{-3.9}$ d | 143.0 | +5.8 | 1060.0 | +8.1 |
| | dSHO | $\mathcal{N}(39,5)$ | $42.8^{+9.1}_{-2.7}$ d | 136.7 | −0.5 | 1056.9 | +4.9 |
| | **dSHO** | $\mathcal{N}(78,5)$ | $76.6^{+4.1}_{-4.2}$ d | **147.2** | **+10.0** | **1064.6** | **+12.6** |

**TABLE C2** Priors and posteriors from the best fit of the 6-planet model. The $t_{\rm conj}$ is equivalent to time in BJD-2 458 354

| | | Posterior | Prior |
|---|---|---|---|
| L 98-59.06 | $K$ [m s$^{-1}$] | $0.70^{+0.14}_{-0.14}$ | $\mathcal{U}(0., 2.0)$ |
| | $M_P \sin i$ [$M_\oplus$] | $0.58^{+0.12}_{-0.12}$ | |
| | $P$ [d] | $1.736\,15^{+0.000\,74}_{-0.000\,76}$ | $\mathcal{U}(1.730, 1.745)$ |
| | $h$ | $-0.02^{+0.13}_{-0.14}$ | trunc$\mathcal{N}(0, 0.15, -0.7, 0.7)$ |
| | $k$ | $0.00^{+0.13}_{-0.13}$ | trunc$\mathcal{N}(0, 0.15, -0.7, 0.7)$ |
| | $e$ | $0.027^{+0.040}_{-0.020}$ | |
| | $\omega$ [°] | $-16^{+110}_{-80}$ | |
| | $a$ [au] | $0.018\,78^{+0.000\,41}_{-0.000\,44}$ | |
| | $\lambda$ [°] | $300^{+22}_{-24}$ | $\mathcal{U}_{\rm cyclic}(85.9, 445.9)$ |
| | $T_{\rm eq}$ [K] | 673 | |
| | $t_{\rm conj}$ [d] | $-1.8^{+0.1}_{-1.6}$ | |
| Planet b | $K$ [m s$^{-1}$] | $0.36^{+0.13}_{-0.13}$ | $\mathcal{U}(0., 1.2)$ |
| | $M_P$ [$M_\oplus$] | $0.32^{+0.12}_{-0.12}$ | |
| | $P$ [d] | $2.253\,114\,19^{+0.000\,000\,17}_{-0.000\,000\,18}$ | $\mathcal{U}(2.2531, 2.2532)$ |
| | $R_p/R_s$ | $0.025\,804^{+0.000\,041}_{-0.000\,041}$ | $\mathcal{U}(0.024, 0.026)$ |
| | $R_p$ [$M_\oplus$] | $0.884^{+0.025}_{-0.025}$ | |
| | $b$ | $0.526^{+0.013}_{-0.013}$ | $\mathcal{U}(0, 1)$ |
| | $h$ | $-0.020^{+0.084}_{-0.094}$ | trunc$\mathcal{N}(0, 0.12, -0.7, 0.7)$ |
| | $k$ | $0.01^{+0.11}_{-0.10}$ | trunc$\mathcal{N}(0, 0.12, -0.7, 0.7)$ |
| | $e$ | $0.015^{+0.018}_{-0.012}$ | |
| | $\omega$ [deg] | $-23^{+130}_{-110}$ | |
| | $a$ [au] | $0.022\,34^{+0.000\,49}_{-0.000\,52}$ | |
| | $a/R_s$ | $15.287\,994\,81^{+0.000\,000\,77}_{-0.000\,000\,81}$ | |
| | $\lambda$ [deg] | $305.3^{+1.1}_{-1.0}$ | |
| | $t_{\rm conj}$ [d] | $2.201\,36^{+0.000\,13}_{-0.000\,13}$ | $\mathcal{U}(0.904, 0.906)$ |
| | $t_{\rm dur}$ [d] | $0.039\,87^{+0.000\,49}_{-0.000\,63}$ | |
| | $T_{\rm eq}$ [K] | 618 | |
| | $\rho_P$ [g cm$^{-3}$] | $2.6^{+1.0}_{-0.9}$ | |

**TABLE C2** Continued.

| | | Posterior | Prior |
|---|---|---|---|
| Planet c | $K$ [m s$^{-1}$] | $2.08^{+0.12}_{-0.12}$ | $\mathcal{U}(0., 3)$ |
| | $M_P$ [$M_\oplus$] | $2.22^{+0.17}_{-0.16}$ | |
| | $P$ [d] | $3.690\,674\,04^{+0.000\,000\,40}_{-0.000\,000\,39}$ | $\mathcal{U}(3.69050, 3.69068)$ |
| | $R_p/R_s$ | $0.039\,60^{+0.000\,31}_{-0.000\,24}$ | $\mathcal{U}(0.0370, 0.0450)$ |
| | $R_p$ [$M_\oplus$] | $1.358^{+0.040}_{-0.040}$ | |
| | $b$ | $0.458^{+0.029}_{-0.021}$ | $\mathcal{U}(0, 1)$ |
| | $h$ | $-0.107^{+0.094}_{-0.085}$ | trunc$\mathcal{N}(0., 0.1, -0.7, 0.7)$ |
| | $k$ | $-0.001^{+0.084}_{-0.084}$ | trunc$\mathcal{N}(0., 0.1, -0.7, 0.7)$ |
| | $e$ | $0.020^{+0.024}_{-0.015}$ | |
| | $\omega$ [deg] | $-82^{+64}_{-43}$ | |
| | $a$ [au] | $0.031\,04^{+0.000\,69}_{-0.000\,72}$ | |
| | $a/R_s$ | $21.243\,679\,9^{+0.000\,001\,5}_{-0.000\,001\,5}$ | |
| | $\lambda$ [deg] | $2352^{+25}_{-25}$ | |
| | $t_{\text{conj}}$ [d] | $2.201\,36^{+0.000\,13}_{-0.000\,13}$ | $\mathcal{U}(2.2005, 2.2200)$ |
| | $t_{\text{dur}}$ [d] | $0.0490^{+0.0008}_{-0.0013}$ | |
| | $T_{\text{eq}}$ [K] | 524 | |
| | $\rho_P$ [g cm$^{-3}$] | $4.88^{+0.60}_{-0.53}$ | |
| Planet d | $K$ [m s$^{-1}$] | $1.39^{+0.13}_{-0.13}$ | $\mathcal{U}(0., 3)$ |
| | $M_P$ [$M_\oplus$] | $1.87^{+0.19}_{-0.19}$ | |
| | $P$ [d] | $7.450\,731^{+0.000\,015}_{-0.000\,015}$ | $\mathcal{U}(7.45060, 7.45085)$ |
| | $R_p/R_s$ | $0.046\,00^{+0.000\,55}_{-0.000\,51}$ | $\mathcal{U}(0.04, 0.05)$ |
| | $R_p$ [$M_\oplus$] | $1.576^{+0.049}_{-0.048}$ | |
| | $b$ | $0.924^{+0.006}_{-0.011}$ | $\mathcal{U}(0, 1)$ |
| | $h$ | $-0.051^{+0.077}_{-0.083}$ | trunc$\mathcal{N}(0., 0.1, -0.7, 0.7)$ |
| | $k$ | $-0.017^{+0.092}_{-0.090}$ | trunc$\mathcal{N}(0., 0.1, -0.7, 0.7)$ |
| | $e$ | $0.013^{+0.016}_{-0.010}$ | |
| | $\omega$ [deg] | $-74^{+143}_{-69}$ | |
| | $a$ [au] | $0.0496^{+0.0011}_{-0.0012}$ | |
| | $a/R_s$ | $33.933\,422\,0^{+0.000\,004\,4}_{-0.000\,004\,5}$ | |
| | $\lambda$ [deg] | $27.9^{+1.0}_{-1.1}$ | |
| | $t_{\text{conj}}$ [d] | $1.289\,03^{+0.000\,22}_{-0.000\,22}$ | $\mathcal{U}(1.270, 1.295)$ |
| | $t_{\text{dur}}$ [d] | $0.025\,94^{+0.000\,53}_{-0.000\,60}$ | |
| | $T_{\text{eq}}$ [K] | 414 | |
| | $\rho_P$ [g cm$^{-3}$] | $2.63^{+0.38}_{-0.34}$ | |

**TABLE C2** Continued.

| | | Posterior | Prior |
|---|---|---|---|
| Planet e | $K$ [m s$^{-1}$] | $1.82^{+0.14}_{-0.14}$ | $\mathcal{U}(0., 3)$ |
| | $M_P \sin i$ [$M_\oplus$] | $2.92^{+0.26}_{-0.26}$ | |
| | $P$ [d] | $12.813^{+0.017}_{-0.017}$ | $\mathcal{U}(12.70, 13.1)$ |
| | $h$ | $0.06^{+0.14}_{-0.13}$ | trunc$\mathcal{N}(0., 0.15, -0.7, 0.7)$ |
| | $k$ | $-0.12^{+0.13}_{-0.12}$ | trunc$\mathcal{N}(0., 0.15, -0.7, 0.7)$ |
| | $e$ | $0.039^{+0.059}_{-0.030}$ | |
| | $\omega$ [deg] | $80^{+70}_{-230}$ | |
| | $a$ [au] | $0.0712^{+0.0016}_{-0.0016}$ | |
| | $\lambda$ [deg] | $231^{+11}_{-11}$ | $\mathcal{U}(0, 360)$ |
| | $T_{eq}$ [K] | 346 | |
| | $t_{conj}$ [d] | $-5^{+1}_{-12}$ | |
| Planet f | $K$ [m s$^{-1}$] | $1.53^{+0.22}_{-0.24}$ | $\mathcal{U}(0., 3)$ |
| | $M_P \sin i$ [$M_\oplus$] | $2.97^{+0.46}_{-0.48}$ | |
| | $P$ [d] | $23.069^{+0.081}_{-0.082}$ | $\mathcal{U}(22.5, 24.5)$ |
| | $h$ | $0.01^{+0.18}_{-0.18}$ | trunc$\mathcal{N}(0., 0.2, -0.7, 0.7)$ |
| | $k$ | $0.18^{+0.15}_{-0.18}$ | trunc$\mathcal{N}(0., 0.2, -0.7, 0.7)$ |
| | $e$ | $0.073^{+0.078}_{-0.055}$ | |
| | $\omega$ [deg] | $4^{+58}_{-64}$ | |
| | $a$ [au] | $0.1053^{+0.0024}_{-0.0024}$ | |
| | $\lambda$ [deg] | $228^{+20}_{-21}$ | $\mathcal{U}(0, 360)$ |
| | $T_{eq}$ [K] | 285 | |
| | $t_{conj}$ [d] | $-9.5^{+1.4}_{-1.6}$ | |
| GP (dSHO) | $\sigma$ | $1.86^{+0.29}_{-0.23}$ | $\ln \mathcal{U}(0.001, 20)$ |
| | $P_{rot}$ | $76.6^{+4.1}_{-4.2}$ | $\mathcal{N}(39, 5)$ |
| | $Q0$ | $1.1^{+1.0}_{-0.6}$ | $\mathcal{U}(0.01, 1000)$ |
| | $dQ$ | $873^{+430}_{-540}$ | $\mathcal{U}(0.01, 1500)$ |
| | $f$ | $18^{+10}_{-10}$ | $\mathcal{U}(0, 35)$ |
| Data Parameters | offset TESS | $-0.000\,036^{+0.000\,002}_{-0.000\,002}$ | $\mathcal{U}(-0.001, 0.001)$ |
| | $u_{1,TESS}$ | $0.335^{+0.073}_{-0.073}$ | $\mathcal{U}(0, 1)$ |
| | $u_{2,TESS}$ | $0.32^{+0.17}_{-0.12}$ | $\mathcal{U}(0, 1)$ |
| | offset HST | $-0.000\,000\,25^{+0.000\,000\,14}_{-0.000\,000\,14}$ | $\mathcal{U}(-0.0005, 0.0005)$ |
| | $u_{1,HST}$ | $0.0468^{+0.0088}_{-0.0080}$ | $\mathcal{U}(0, 1)$ |
| | $u_{2,HST}$ | $0.006^{+0.010}_{-0.005}$ | $\mathcal{U}(0, 1)$ |
| | offset HARPS | $1.69^{+0.75}_{-0.76}$ | $\mathcal{U}(-10, 10)$ |
| | log(Jitter) HARPS | $0.04^{+0.22}_{-0.03}$ | $\mathcal{U}(-8, 3)$ |
| | offset $\text{ESPRESSO}_{pre}$ | $1.95^{+0.86}_{-0.90}$ | $\mathcal{U}(-10, 10)$ |
| | log(Jitter) $\text{ESPRESSO}_{pre}$ | $0.05^{+0.34}_{-0.05}$ | $\mathcal{U}(-8, 3)$ |
| | offset $\text{ESPRESSO}_{post}$ | $3.3^{+2.2}_{-2.2}$ | $\mathcal{U}(-10, 10)$ |
| | log(Jitter) $\text{ESPRESSO}_{post}$ | $0.04^{+0.25}_{-0.04}$ | $\mathcal{U}(-8, 3)$ |
| | RV linear coefficient | $-0.0072^{+0.0044}_{-0.0043}$ | $\mathcal{U}(-0.1, 0.01)$ |